\documentclass[%
 aip,
 amsmath,amssymb,
reprint,%
]{revtex4-1}

\usepackage{graphicx}
\usepackage{dcolumn}
\usepackage{bm}

\usepackage[utf8]{inputenc}
\usepackage[T1]{fontenc}
\usepackage{mathptmx}
\usepackage{textcomp}

\usepackage{color}

\begin{document}

\preprint{AIP/123-QED}

\title[Two-dimensional hybrid model of gradient drift instability and enhanced electron transport in a Hall thruster]{Two-dimensional hybrid model of gradient drift instability and enhanced electron transport in a Hall thruster}

\author{R. Kawashima}
 \email{kawashima@al.t.u-tokyo.ac.jp.}
\author{K. Komurasaki}%
\affiliation{ 
Department of Aeronautics and Astronautics, The University of Tokyo, Tokyo 113-8656, Japan
}%


\date{\today}

\begin{abstract}
An axial-azimuthal two-dimensional Hall thruster discharge model was developed for analyzing gradient drift instability (GDI) and cross-field electron transport enhancement induced solely by the GDI.
A hybrid particle-fluid model was used for the partially ionized plasma, where the inertialess electron fluid in the quasineutral plasma was assumed.
A nonoscillatory numerical method was proposed for the potential solver in the electron fluid model to avoid numerical instability and analyze the physics of GDI accurately.
A simulation is performed for a 1 kW-class anode-layer-type Hall thruster, and the flow field with plasma instability is presented.
Plasma instability with vortex-like structures is observed in the acceleration and plume regions.
The generated plasma instability enhances the cross-field electron transport in the axial direction around the channel exit and in the plume region.
Grid convergence is confirmed regarding the effect of electron transport enhancement, which indicates that cross-field electron transport enhancement is based on the plasma instability.
Furthermore, the comparison between the simulation results and linear perturbation analyses demonstrates that the simulated plasma instability reflects the theory of GDI.
Thus, it is concluded that the hybrid model is useful for the analyses of GDI, and the GDI can enhance the cross-field electron transport in Hall thrusters.
\end{abstract}

\maketitle

\section{\label{sec:intro}Introduction}

Numerical modeling technology of partially ionized plasma flow has advanced, and computer-aided-engineering (CAE) is now used in the development processes of plasma sources.
A Hall thruster is an efficient electron propulsion system with a cross-field configuration of the axial electric field and radial magnetic field.
Hall thrusters essentially have two variants: a stationary plasma thruster (SPT) and a thruster with anode layer (TAL) \cite{Choueiri:2001aa}.
In comparison to the SPT, the TAL is characterized by a short discharge channel with a length of a few millimeters and conducting channel walls biased at the cathode potential.
Owing to recent experimental studies on design optimization, a 5 kW-class TAL attained high performance that was competitive with the performance of SPT-type thrusters at the same power level \cite{HamadaIEPC2017}.
A high-fidelity plasma flow simulation is essential to develop Hall thrusters, including SPT and TAL, and to further improve performance.

A challenge in the numerical modeling of Hall thruster discharge is the self-consistent modeling of anomalous electron transport.
It is well known that discharge current and plasma property distributions observed in experiments cannot be reproduced by numerical simulations if one assumes only the classical diffusion theory for the cross-field electron transport.
Further, in the axisymmetric modeling of Hall thrusters, a numerical solution of stable plasma discharge cannot be obtained with only the classical diffusion \cite{Boeuf:2017aa}.
Therefore, it is necessary to add an ad-hoc function for the ``anomalous'' part in the electron transport to match the numerical simulation results with measurements.
For SPTs, there have been several studies on the empirical expression for the anomalous electron transport \cite{KooPoP2006,Hagelaar:2002aa}.
However, for TALs, experimental data to obtain the anomalous electron transport property are lacking, and an empirical model is not available.
Because the channel wall and magnetic field configuration of TAL is significantly different from that of an SPT, the electron transport property in a TAL is hardly predictable based on the knowledge of SPTs.
In fact, the issue of anomalous electron transport has been found in other plasma sources with E$\times$B configurations such as the Penning discharge and magnetron sputtering devices, and this issue entails a general scientific concern \cite{Keidar:2006aa}.
A self-consistent (nonempirical) model is desired for characterizing anomalous electron transport to achieve further high-fidelity numerical simulations for SPT, TAL, and other E$\times$B plasma sources.


One hypothesis for the mechanism of anomalous electron transport is associated with plasma instability that develops in the cross-magnetic field (E and E$\times$B) directions.
According to recent experimental and theoretical studies, anomalous electron transport is attributed to two types of plasma instabilities: electron cyclotron drift instability (ECDI) and gradient drift instability (GDI).

The ECDI is induced by the difference in the flow velocities of the ions and electrons, and hence, this instability is supposed to be the Kelvin--Helmholtz type \cite{Lucken:2019aa}.
In Hall thrusters, because the difference in the flow velocity originates from the azimuthal flow velocity of the electron, the ECDI is generated around the channel exit and plume regions where the Hall parameter is high.
The characteristics of ECDI have been investigated using kinetic particle-in-cell (PIC) simulations because this instability is induced by kinetic electron flow.
A pioneering research of the PIC simulation for ECDI was achieved by Hirakawa \cite{Hirakawa:1996aa,HirakawaIEPC1997} for the radial-azimuthal (R-$\theta$) plane of the Hall thruster.
Several particle simulations were performed for the axial-azimuthal (Z-$\theta$) coordinate, and coherent plasma structures propagating in the azimuthal direction were observed \cite{Adam:2003aa,Lafleur:2016aa,Katz:2018aa}.
In addition, it was confirmed that the ECDI increased the cross-field electron transport, and its contribution was greater than that of the classical diffusion theory \cite{Lafleur:2017aa}.

The GDI is caused by gradients of plasma density and magnetic flux density; therefore, the GDI belongs to the Rayleigh--Taylor type \cite{Kapulkin:2008aa}.
This instability can also be considered as a collisionless Simon--Hoh instability \cite{Simon:1963ux,Hoh:1963vi}.
In Hall thrusters, the GDI is believed to be induced in the acceleration region, where the axial gradients of the plasma and magnetic flux densities are both negative toward the downstream.
This region practically coincides with the region where the ECDI arises, and the mixed ECDI and GDI can be generated in the acceleration region.
The characteristics of the GDI in the Hall thruster discharge have been investigated by perturbation theories, including linear models \cite{Esipchuk:1976,FriasPoP2012}, nonlinear models with the electron inertia and gyroviscosity effects \cite{Smolyakov:2016aa}, and global stability analyses \cite{Escobar2015IEPC}.
Most of these models are derived from the fluid equations of ions and electrons.
Perturbation models are powerful tools for analyzing the characteristics of the GDI, such as the frequency and growth rate.

However, numerical plasma flow modeling is useful for simulating the interaction between the steady-state solution and the instabilities.
Several numerical simulations related to the GDI have been conducted for the Z-$\theta$ coordinate of the Hall thruster.
Lam et al. performed a hybrid particle-fluid simulation, where complex plasma fluctuations were induced in both $\pm$E$\times$B directions \cite{Lam6922570}.
Similar hybrid kinetic-fluid simulations were performed by Fernandez et al. \cite{Fernandez2015iepc} and Hara et al. \cite{HaraIEPC2015}, and they reported the issue of numerical instability in the potential calculations.

A conclusive physical explanation for the effect of GDI on electron transport is lacking.
The abovementioned numerical study mentioned that the effect of electron transport enhancement by the GDI was insignificant compared with the transport properties observed in the experiments \cite{LamPhDThesis}.
However, the fully kinetic simulation in Ref. \onlinecite{Chernyshev:2019aa} mentioned that the cross-field electron transport is strongly affected by the GDI in addition to ECDI.
Further detailed numerical simulations are required to clarify the characteristics of GDI and its effects on cross-field electron transport.
Based on previous numerical simulation studies, a hybrid model with nonoscillatory numerical methods should be developed for the analysis of the GDI, and it should be carefully examined if the developed model yields the enhanced cross-field electron transport.

The objective of this study is to develop a hybrid particle-fluid simulation that resolves the GDI on the Z-$\theta$ coordinate and examine if the simulation yields enhanced cross-field electron transport.
Unlike the fully particle model, the electron fluid model used in the hybrid simulation does not yield the kinetic ECDI.
Therefore, electron transport enhancement induced solely by the GDI is modeled using a hybrid model.
In Ref. \onlinecite{Kawashima:2018ab}, we performed a hybrid simulation for analyzing the rotating spokes in an SPT assuming the empirical cross-field electron transport property \cite{Kawashima:2018ab}.
In this study, the hybrid model is applied to a TAL, with the electron transport model based only on classical diffusion theory.
A nonoscillatory scheme was developed for the magnetized electron fluid calculation to avoid the numerical instability arising in the potential calculation, and therefore, the simulation reflects the physics of the GDI.
The effects of the instability on the cross-field electron transport were examined through the analyses of the effective electron mobility and Hall parameters.

\section{Axial-Azimuthal Hybrid Model}

\begin{figure}
    \includegraphics[width=80mm]{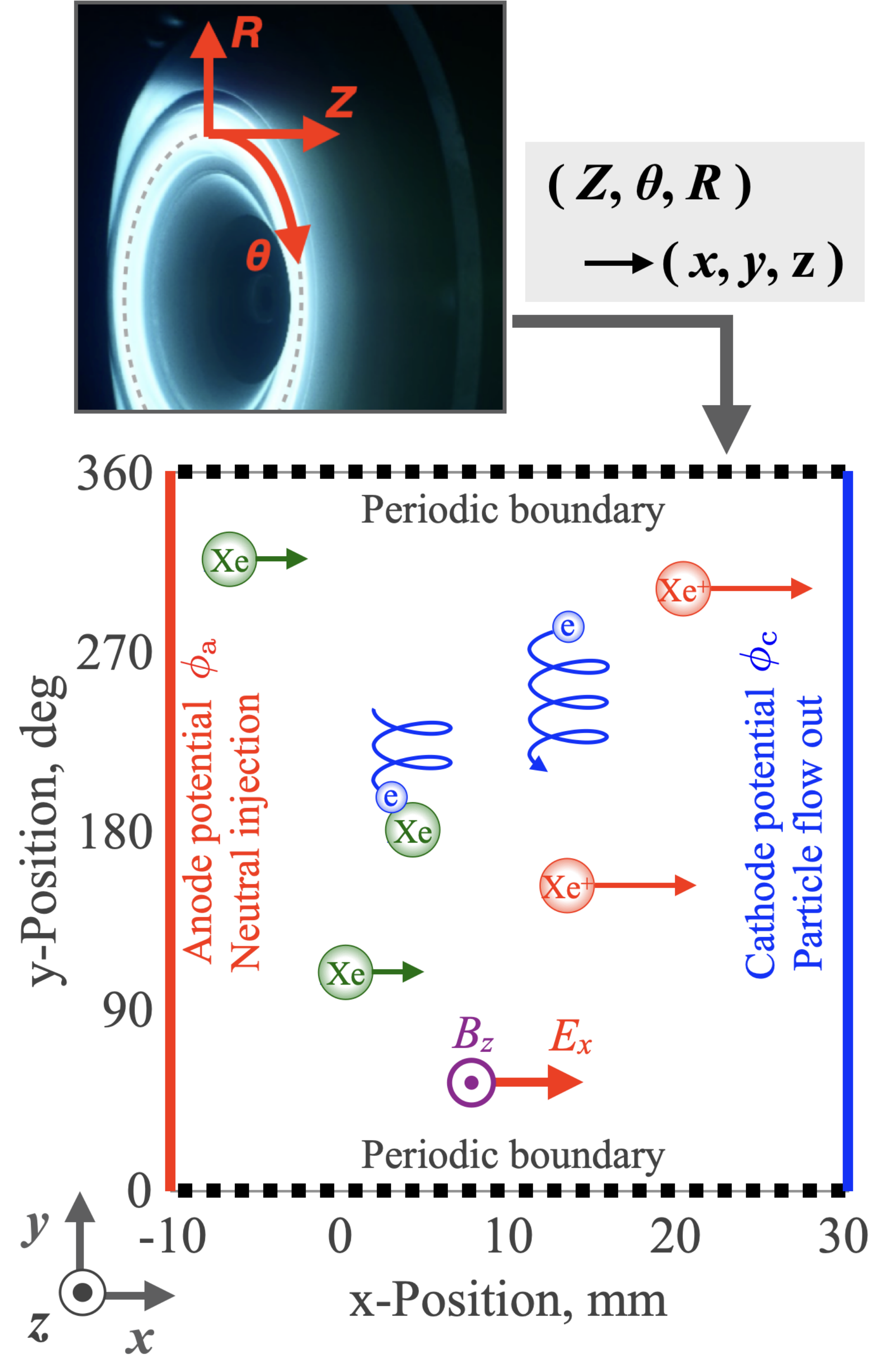}
    \caption{\label{fig:condition} Simulation domain for the two-dimensional model that corresponds to the Z-$\theta$ coordinate in the Hall thruster.}
\end{figure} 

The simulation setup for the Z-$\theta$ two-dimensional model is shown in Fig. \ref{fig:condition}.
The coordinate system was converted from $\left(Z, \theta, R\right)$ to $\left(x, y, z\right)$, neglecting the effects of the curvature of the annular discharge channel. The x- and y-axes correspond to the axial and azimuthal directions of the discharge channel, respectively.
A quasineutral particle-fluid hybrid model is used wherein ions and neutral particles are modeled as particles, and electrons are approximated as a fluid in the plasma.
In the numerical simulations of the Hall thrusters, the hybrid model has been used because of the fidelity of the modeling and the reasonable computational cost \cite{FifeThesis,Komurasaki:1995fk}.
Compared to the fully kinetic models, the major assumptions considered in the present hybrid model are:
\begin{enumerate}
\renewcommand{\labelenumi}{\arabic{enumi})}
\setlength{\leftskip}{5.0mm}
    \item Maxwellian electron velocity distribution function (fluid approximation),
    \item quasineutrality (plasma approximation)
    \item inertialess electrons (drift-diffusion model)
\end{enumerate}
As explained in Ref. \onlinecite{Kawashima:2018ab}, this type of hybrid model is advantageous for long-time simulations of plasma flow in large domains.
In this study, the basics of this model are explained briefly.

\subsection{Ion and neutral particles}
Ion and neutral particle flows are calculated using the PIC method.
A cloud of $\sim10^8$ particles is treated as one macroparticle, and the motion of each macroparticle is tracked by calculating the equation of motion.
The effects of ion magnetization are neglected for simplicity. Thus, the ions are accelerated only electrostatically.
The time integration for the equation of motion is implemented via a second-order leapfrog scheme.
The ionization and electron-neutral scattering collision frequencies are calculated using reaction rate coefficients, which are provided as empirical functions of electron temperature \cite{GoebelKatz2008}.
Neutral particles are injected from the anode-side boundary with velocity distributions based on the cosine law \cite{Greenwood:2002aa}.
These neutral particles are ionized in the calculation domain and ejected from the cathode-side boundary.

\subsection{Electron fluid}

The electron fluid model assumes the inertialess electrons in a quasineutral plasma.
The mass conservation equation is
\begin{equation}
    \nabla\cdot\left(n_{\rm e}\vec{u}_{\rm e}\right)
    =S_{\rm ion},
    \label{eq:mass}
\end{equation}
where $n_{\rm e}$, $u_{\rm e}$, and $S_{\rm ion}$ denote the electron number density, electron velocity, and ionization rate, respectively.
It is assumed that electrons are sufficiently mobile to achieve quasineutrality instantaneously, and the time-derivative term is omitted from the equation.
Further, $S_{\rm ion}$ is the source term of ionization, and the same amount of ions and electrons are generated in the PIC and electron fluid parts, respectively.

Electron momentum conservation is written in the form of the drift-diffusion equation.
In the x-y plane with magnetic fluxes in the z-direction, the equation is written as
\begin{equation}
   \mu_\perp
    \left[\begin{array}{cc}1 & \Omega_{\rm e} \\
    -\Omega_{\rm e} & 1\end{array}\right]\Bigl(n_{\rm e}\nabla\phi
    -\nabla\left(n_{\rm e}T_{\rm e}\right)\Bigr)
    =n_{\rm e}\vec{u}_{\rm e},
    \label{eq:momentum}
\end{equation}
where $\phi$, $T_{\rm e}$, and $\mu_\perp$ denote the plasma potential, electron temperature, and cross-field electron mobility, respectively.
The electron Hall parameter $\Omega_{\rm e}$ is
\begin{equation}
    \Omega_{\rm e}=\frac{eB_z}{m_{\rm e}\nu_{\rm col}},
    \label{eq:Omega}
\end{equation}
where $e$, $B$, $m_{\rm e}$, and $\nu_{\rm col}$ denote the elementary charge, magnetic flux density, electron mass, and total collision frequency, respectively.
Further, $\nu_{\rm col}$ is simply modeled using the empirical functions of the electron--neutral scattering collision frequency \cite{GoebelKatz2008}.
The electron--ion Coulomb collision is neglected.
The cross-field electron mobility is expressed via the classical diffusion theory as
\begin{equation}
    \mu_\perp=\frac{1}{1+\Omega_{\rm e}^2}\cdot\frac{e}{m_{\rm e}\nu_{\rm col}}.
    \label{eq:muperp}
\end{equation}
The cross-field electron mobility expressed in Eq. (\ref{eq:muperp}) is called the classical electron mobility.
No artificial collision frequency or electron mobility model is used to account for the anomalous electron transport, because this simulation aims to analyze the anomalous electron transport induced by the GDI.

The electron energy conservation is described for the electron internal energy as
\begin{align}
    \frac{\partial}{\partial t}\left(\frac{3}{2}n_{\rm e}T_{\rm e}\right)
    +\nabla\cdot\left(\frac{5}{2}\vec{\Gamma}_{\rm e,ave}T_{\rm e}
    -\kappa\nabla T_{\rm e}\right)&\nonumber\\
    =\vec{\Gamma}_{\rm e,ave}\cdot\nabla\phi
    -\alpha_{\rm E}&\varepsilon_{\rm ion}S_{\rm ion},
    \label{eq:ene}
\end{align}
where $\Gamma_{\rm e}$, $\kappa$, and $\varepsilon_{\rm ion}$ are the electron number flux, coefficient for the heat conductivity, and the ionization potential, respectively.
$\alpha_{\rm E}$ is a dimensionless coefficient accounting for energy losses caused by inelastic collisions and radiation, and it is modeled as
$\alpha_{\rm E}= 2.0 + 0.254 \exp\left(2\varepsilon_{\rm ion}/3T_{\rm e}\right)$ \cite{Dugan_Sovie_1967}.
$\Gamma_{\rm e,ave}$ is the azimuthally averaged electron flux.
As shown in Sec. \ref{sec:dist}, electron flow is highly oscillatory, and the numerical simulation occasionally becomes unstable if the two-dimensional $\Gamma_{\rm e}$ is used in the electron energy equation.
Hence, the azimuthal (y-direction) average is used for $\Gamma_{{\rm e},x}$ and $\Gamma_{{\rm e},y}$ for simplicity.
In the previous hybrid models for the axial-azimuthal simulations, the azimuthal average was used for all quantities in the electron energy equation \cite{Kawashima:2018ab,Lam6922570}.
In the present model, only $\vec{\Gamma}_{\rm e}$ is azimuthally averaged, and all other quantities such as $\phi$ and $T_{\rm e}$ are treated in two dimensions in Eq. (\ref{eq:ene}).
This model continues to yield a solution of electron temperature that is essentially one-dimensional in the x-(axial) direction.
Hence, the effects of plasma instability on electron energy flow are partially included in the present model.

The coefficient for heat conduction is modeled as a scalar quantity as
\begin{equation}
    \kappa=\frac{5}{2}n_{\rm e}T_{\rm e}\mu_\perp.
\end{equation}
The use of $\mu_\perp$ corresponds to the diffusion coefficient of the magnetized electrons.
In the axial-radial models for Hall thrusters, the coefficient for heat conduction is modeled in a tensor form to account for anisotropic heat conduction in the parallel and perpendicular directions of the magnetic lines of force \cite{KawashimaIEPC2019}.
In the present model for the axial-azimuthal coordinate, the heat conductivity is modeled as an isotropic scalar quantity.
The heat fluxes carried by the azimuthal Hall current are included in the convection term in Eq. (\ref{eq:ene}).

\subsection{Wall effect}
\begin{figure}
    \includegraphics[width=80mm]{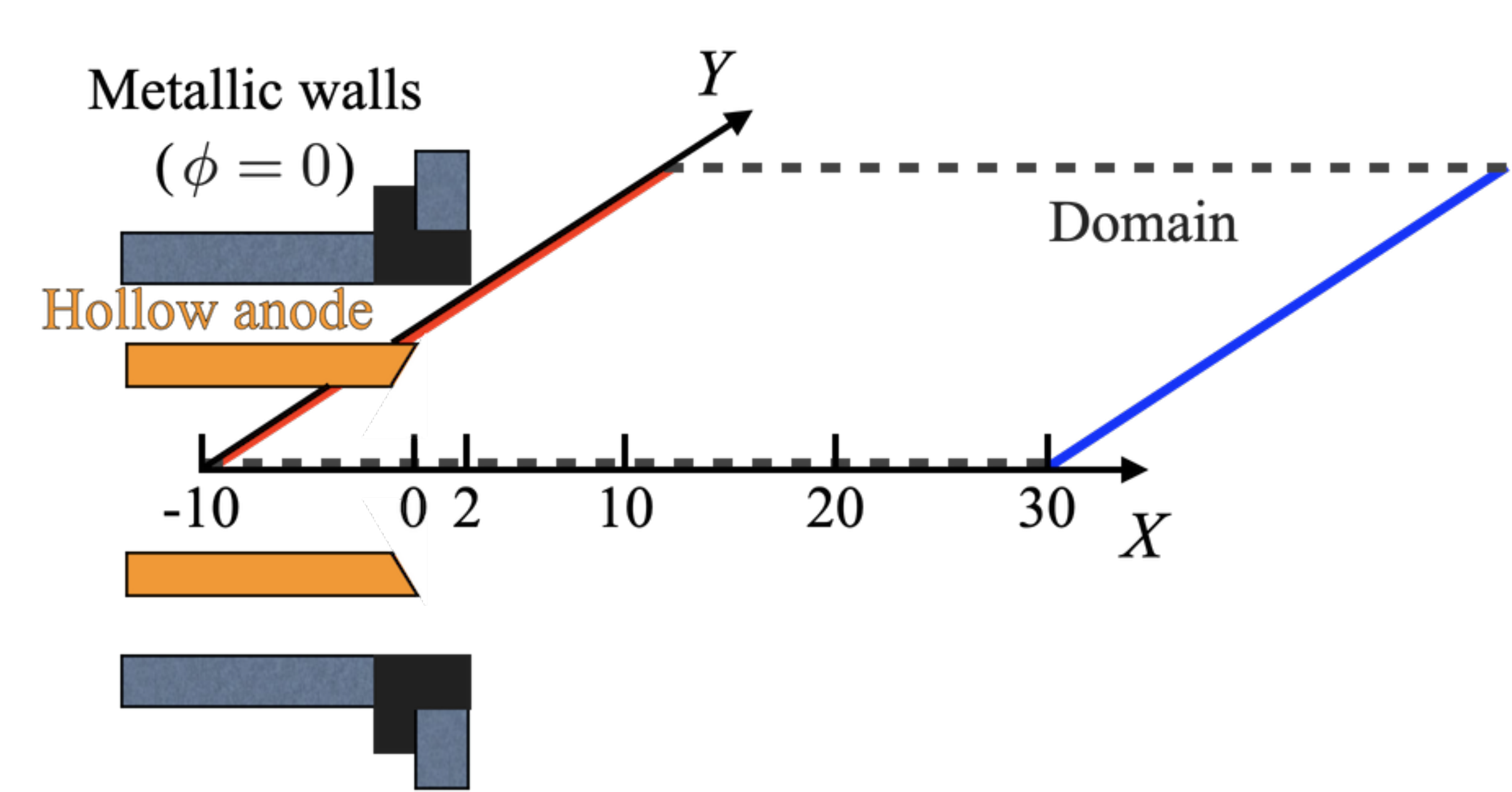}
    \caption{\label{fig:wall} Relationship between the calculation domain and thruster geometry.
    $x=0$ is defined at the hollow anode tip and the channel exit is at $x=2$ mm.}
\end{figure}

The TAL-type Hall thruster uses metallic channel walls and a hollow anode as the lateral wall.
The relationship between the simulation domain and the walls is shown in Fig. \ref{fig:wall}.
The ion flux lost to the inner and outer channel walls and the hollow anode is modeled using the electrostatic sheath theory, as \cite{Lam6922570}
\begin{equation}
    \Gamma_{\rm wall}=2\times n_{\rm e}\exp\left(-0.5\right)\sqrt{\frac{eT_{\rm e}}{m_{\rm i}}}.
    \label{eq:wall}
\end{equation}
The volumetric source term for the ion wall loss is written as $S_{\rm wall}= \Gamma_{\rm wall}/w$, where $w$ denotes the width of the discharge channel or hollow anode.
In each computational grid, a fraction of ions is changed to neutral particles corresponding to the ion wall recombination of $S_{\rm wall}$.
Neutral particles generated in this process have initial velocities of the Maxwellian distribution function of the wall temperature.

It is assumed that the electrons are not lost to the inner and outer channel walls.
In TALs, the metallic channel walls are biased at the cathode potential, and the sheath in front of the walls repels most electrons.
Thus, the electron and electron energy fluxes flowing to the channel walls are assumed to be negligibly small.
Electrons flow into the hollow anode from the discharge plasma.
This means that electrons contained in the plasma near the hollow anode smoothly flow out from the simulation domain.
To reflect the effect of the hollow anode in the present simulation, electron mobility is artificially increased in the region included in the hollow anode (-10 mm $\leq x < 0$ mm).
That is, the collision frequency inside the hollow anode is changed to $\nu_{\rm col}\rightarrow \left(1+\alpha_{\rm HA}\right)\nu_{\rm col}$.
However, because magnetic confinement occurs around the channel exit and plume region, the effects of the collision frequency and electron conductivity in the hollow anode region are minor.
We executed the simulation with a hollow anode parameter $\alpha_{\rm HA}$ of 1, 10, and 20.
In these simulation results, the plasma potential in the hollow anode region was almost uniform, and the difference in the discharge current was insignificant.
Thus, simulation results obtained with $\alpha_{\rm HA}$ of 1 are presented in this paper.

\section{Numerical method for electron fluid}
\subsection{Potential solver}
\label{sec:pot}
In the electron fluid model with the quasineutraility assumption, the plasma potential is obtained from electron fluid equations but not from Poisson's equation.
The conservation equations of mass in Eq. (\ref{eq:mass}) and momentum in Eq. (\ref{eq:momentum}) were utilized to calculate the plasma potential.
There are two numerical approaches: the elliptic-equation (EE) approach and the hyperbolic-system (HS) approach.
The EE approach uses a diffusion equation obtained by inserting Eq. (\ref{eq:momentum}) into Eq. (\ref{eq:mass}).
This approach is a common method for calculating the plasma potential in a two-dimensional hybrid model for Hall thrusters \cite{Komurasaki:1995fk,Hagelaar:2007aa,Lam6922570}.
The HS approach uses a hyperbolic-equation system obtained by adding the pseudo-time advancement terms to Eqs. (\ref{eq:mass}) and (\ref{eq:momentum}).
This approach was developed by the authors for the robust computation of magnetized electron fluid \cite{Kawashima201559,Kawashima2016202}.
Further, the HS approach was used to calculate the axial-azimuthal electron fluid model for Hall thrusters \cite{Kawashima:2018ab,Kawashima:2018aa} in a stable manner.
In this study, by using the knowledge of the HS approach, a stable numerical method for the EE approach is reconsidered.

The EE approach handles the second-order elliptic equation obtained from Eqs. (\ref{eq:mass}) and (\ref{eq:momentum}) as
\begin{align}
    \nabla\cdot&\Bigl(
    -n_{\rm e}\mu_\perp\left[\begin{array}{cc}1 & \Omega_{\rm e} \\
    -\Omega_{\rm e} & 1\end{array}\right]\nabla\phi\Bigr)
    \nonumber \\
    =&\nabla\cdot\Bigl(
    -\mu_\perp\left[\begin{array}{cc}1 & \Omega_{\rm e} \\
    -\Omega_{\rm e} & 1\end{array}\right]\nabla\left(n_{\rm e}T_{\rm e}\right)
    \Bigr)-S_{\rm ion}.
    \label{eq:pot1}
\end{align}
This equation is solved for $\phi$ as a boundary value problem.
First, the isotropic diffusion and drift terms are separated as
\begin{align}
    \mu_\perp\left[\begin{array}{cc}1 & \Omega_{\rm e} \\
    -\Omega_{\rm e} & 1\end{array}\right]&=\mu_\perp\left[
    \begin{array}{cc}
        1 &  \\
         & 1
    \end{array}
    \right]+\mu_\perp\left[
    \begin{array}{cc}
         & \Omega_{\rm e} \\
        -\Omega_{\rm e} & 
    \end{array}
    \right]
    \nonumber \\
    &=\left[\mu\right]_{\rm diffusion}+\left[\mu\right]_{\rm drift}.
    \label{eq:pot2}
\end{align}
The diffusion terms with $\left[\mu\right]_{\rm diffusion}$ are isotropic diffusion, and simple second-order central differences can be used for these terms.
The discretization method is the same as that commonly used for Poisson's equation solvers; thus, it is not described here.

The discretization method for the drift terms with $\left[\mu\right]_{\rm drift}$ is considered.
This term is a so-called anisotropic diffusion, and it is known that the computation becomes unstable if central differencing is simply applied to this term \cite{Kawashima:2018aa}.
Instead, we propose an approach that converts anisotropic diffusion terms into first-order space-derivative terms.
For instance, the drift term for $\phi$ can be mathematically modified as
\begin{align}
    \nabla\cdot\Bigl(
    -n_{\rm e}\mu_\perp&\left[
    \begin{array}{cc}
         & \Omega_{\rm e} \\
        -\Omega_{\rm e} & 
    \end{array}
    \right]\nabla\phi\Bigr)
    \nonumber \\
    &=\nabla\cdot\Bigl(
    -\sigma_\wedge\frac{\partial \phi}{\partial y}
    , \sigma_\wedge\frac{\partial \phi}{\partial x}\Bigr)^T
    \nonumber \\
    &=\nabla\cdot\left(\left[
    \begin{array}{cc}
         & \phi \\
        -\phi & 
    \end{array}
    \right]\nabla\sigma_\wedge\right),
    \label{eq:pot3}
\end{align}
where $\sigma_\wedge\equiv n_{\rm e}\mu_\perp\Omega_{\rm e}$.
In this process, the second-order space derivative is switched from $\phi$ to $\sigma_\wedge$, and the anisotropic diffusion for $\phi$ is avoided.
This process is called differential operator switching (DOS).
In the present simulation, DOS is the key to achieving stable plasma potential computation.

After applying DOS, the term is discretized in the form of the finite volume method as
\begin{align}
    &\nabla\cdot\left(\left[
    \begin{array}{cc}
         & \phi \\
        -\phi & 
    \end{array}
    \right]\nabla\sigma_\wedge\right)\nonumber\\
    &=\frac{1}{\Delta x}\left(F_{x,i+\frac{1}{2},j}-F_{x,i-\frac{1}{2},j}\right)+\frac{1}{\Delta y}\left(F_{y,i,j+\frac{1}{2}}-F_{y,i,j-\frac{1}{2}}\right).
    \label{eq:pot4}
\end{align}
The numerical flux $F$ in the x-direction at the cell boundary is regarded as an advection term for $\phi$; thus, it is evaluated by the upwind method as
\begin{align}
    F_{x,i+\frac{1}{2},j}&=\left(\frac{\partial \sigma_\wedge}{\partial y}\cdot\phi\right)_{i+\frac{1}{2},j}\nonumber\\
    &=\frac{1}{2}\left[\left(\frac{\partial \sigma_\wedge}{\partial y}\right)_{i,j}\cdot\phi_{i,j}
    +\left(\frac{\partial \sigma_\wedge}{\partial y}\right)_{i+1,j}\cdot\phi_{i+1,j}\right]\nonumber\\
    &-\frac{1}{2}\left|\frac{\partial \sigma_\wedge}{\partial y}\right|_{i+\frac{1}{2},j}\left(\phi_{i+1,j}-\phi_{i,j}\right).
    \label{eq:pot5}
\end{align}
The second term on the right-hand side of Eq. (\ref{eq:pot5}) represents numerical diffusion to avoid numerical instabilities associated with this advection term.
The coefficient for the numerical diffusion term at the cell interfaces is evaluated as
\begin{equation}
    \left|\frac{\partial \sigma_\wedge}{\partial y}\right|_{i+\frac{1}{2},j}
    ={\rm max}\left(\left|\frac{\partial \sigma_\wedge}{\partial y}\right|_{i,j},\left|\frac{\partial \sigma_\wedge}{\partial y}\right|_{i+1,j}\right).
    \label{eq:pot6}
\end{equation}
With this evaluation method, the numerical diffusion term in Eq. (\ref{eq:pot5}) becomes sufficiently large for stable computation.
In Eqs. (\ref{eq:pot5}) and (\ref{eq:pot6}), the gradient of $\sigma_\wedge$ is evaluated by applying central differencing as
\begin{equation}
    \left(\frac{\partial \sigma_\wedge}{\partial y}\right)_{i,j}
    =\frac{1}{2\Delta y}\left(\sigma_{\wedge,i,j+1}-\sigma_{\wedge,i,j-1}\right).
    \label{eq:pot7}
\end{equation}
The numerical flux $F$ in the y-direction is evaluated in a similar manner.

In the above explanation, we discussed the discretization method for the diffusion term of the potential. 
The diffusion term of the pressure on the right-hand side of Eq. (\ref{eq:pot1}) is discretized using the same method.
However, for the pressure diffusion term, the numerical diffusion in Eq. (\ref{eq:pot5}) is not used because it does not affect the computational stability of the potential solver.

As written in Eq. (\ref{eq:pot5}), the numerical flux for the drift component at the cell boundary $i+1/2$ is provided with regard to $\phi_{i,j}$ and $\phi_{i+1,j}$.
The numerical flux for the isotropic diffusion term is written with $\phi_{i,j}$ and $\phi_{i+1,j}$, if one applies the standard central differencing.
Therefore, the left-hand side of Eq. (\ref{eq:pot1}) is written in a linear algebraic form with coefficients for $\phi$ at five stencils ($\phi_{i+1,j}$, $\phi_{i,j+1}$, $\phi_{i,j}$, $\phi_{i-1,j}$, and $\phi_{i,j-1}$).
The linear algebraic equation can be solved by direct or iterative methods.
In the present simulation, a direct matrix inversion method based on Gaussian elimination is used to ensure the strict electron flux conservation and minimal residuals.

\subsection{Electron flux calculation}
Electron flux at the cell interfaces must be calculated in a manner consistent with the potential solver.
The electron flux $\Gamma_{\rm e}$ can be divided into the diffusion and drift components as
\begin{align}
    \vec{\Gamma}_{\rm e}
    &=\vec{\Gamma}_{\rm e,diffusion}+\vec{\Gamma}_{\rm e,drift}\nonumber\\
    &=n_{\rm e}\left[\mu\right]_{\rm diffusion}\nabla\phi+n_{\rm e}\left[\mu\right]_{\rm drift}\nabla\phi.
    \label{eq:flux1}
\end{align}
The evaluation method for the diffusion component is the same as the standard central difference method, and thus, the explanation is omitted here.

The drift component of the electron flux is expressed as
\begin{equation}
    \vec{\Gamma}_{{\rm e},{\rm drift}}=\left[\mu\right]_{\rm drift}\Bigl(n_{\rm e}\nabla\phi
    -\nabla\left(n_{\rm e}T_{\rm e}\right)\Bigr).
    \label{eq:flux2}
\end{equation}
Here, we consider the term with $\phi$ for the convenience of explanation. 
The drift component of the electron flux in the x-direction can be converted as
\begin{equation}
    \Gamma_{{\rm e},x,{\rm drift}}=\sigma_\wedge\frac{\partial \phi}{\partial y}=\frac{\partial}{\partial y}\left(\sigma_\wedge \phi\right)-\frac{\partial \sigma_\wedge}{\partial y}\phi.
    \label{eq:flux2}
\end{equation}
This conversion process corresponds to DOS in Eq. (\ref{eq:pot3}).
The electron flux at the cell interface $i+1/2$ is evaluated as
\begin{align}
    \Gamma_{{\rm e},x,{\rm drift},i+\frac{1}{2},j}
    &=\frac{1}{2}\left(\left.\frac{\partial}{\partial y}\left(\sigma_\wedge \phi\right)\right|_{i,j}+\left.\frac{\partial}{\partial y}\left(\sigma_\wedge \phi\right)\right|_{i+1,j}\right)\nonumber\\
    &-\frac{1}{2}\left[\left(\frac{\partial \sigma_\wedge}{\partial y}\right)_{i,j}\phi_{i,j}+\left(\frac{\partial \sigma_\wedge}{\partial y}\right)_{i+1,j}\phi_{i+1,j}\right]\nonumber\\
    &+\frac{1}{2}\left|\frac{\partial \sigma_\wedge}{\partial y}\right|_{i+\frac{1}{2},j}\left(\phi_{i+1,j}-\phi_{i,j}\right).
    \label{eq:flux3}
\end{align}
The evaluation method for the coefficient of the numerical diffusion term (the third term on the right-hand side) is the same as that in Eq. (\ref{eq:pot6}).
Gradients of $\sigma_\wedge \phi$ and $\sigma_\wedge$ are evaluated by the central differences in Eq. (\ref{eq:pot7}).
The drift component of the electron flux in the y-direction was also evaluated in a similar manner.
The expression for the electron flux in Eq. (\ref{eq:flux3}) is different from the numerical flux $F_{x,i+\frac{1}{2},j}$ in Eq. (\ref{eq:pot5}).
As explained in Sec. \ref{sec:transport}, the electron flux calculated by Eq. (\ref{eq:flux3}) strictly conserves the electron flux in the calculation domain.

\subsection{Electron temperature solver}

Electron energy conservation in Eq. (\ref{eq:ene}) is calculated as a time-dependent problem.
The discretization is implemented in the form of a finite volume method.
The heat flux $Q$ in the x-direction at the cell interface is expressed as
\begin{equation}
    Q_{x,i+\frac{1}{2},j}=\left(\frac{5}{2}\Gamma_{{\rm e,ave},x}T_{\rm e}-\kappa\frac{\partial T_{\rm e}}{\partial x}\right)_{i+\frac{1}{2},j}.
    \label{eq:ene2}
\end{equation}
The convection term is evaluated by using the upwind method.
The electron fluxes were originally evaluated at the cell interfaces as shown in Eq. (\ref{eq:flux3}); hence, no interpolation is required.
To interpolate $T_{\rm e}$ at the cell interfaces, the monotonic upstream-centered scheme for conservation laws (MUSCL) technique is used to obtain the second-order spatial accuracy.
The diffusion term for heat conduction is evaluated using second-order central differencing.

The source term in Eq. (\ref{eq:ene}) was evaluated at the cell centers.
The Joule heating term is calculated using the splitting method as
\begin{align}
    \vec{\Gamma}_{\rm e, ave}& \cdot\nabla\phi \nonumber\\
    =&\frac{1}{2}\left[\left(\Gamma_{{\rm e, ave},x}\frac{\partial \phi}{\partial x}\right)_{i+\frac{1}{2},j}+\left(\Gamma_{{\rm e, ave},x}\frac{\partial \phi}{\partial x}\right)_{i-\frac{1}{2},j}\right]\nonumber\\
    +&\frac{1}{2}\left[\left(\Gamma_{{\rm e, ave},y}\frac{\partial \phi}{\partial y}\right)_{i,j+\frac{1}{2}}+\left(\Gamma_{{\rm e, ave},y}\frac{\partial \phi}{\partial y}\right)_{i,j-\frac{1}{2}}\right],
    \label{eq:ene3}
\end{align}
where the gradients of $\phi$ at the cell interfaces are evaluated by the central differences between the two neighboring cells.

Time advancement is implemented using the fully implicit method based on the delta-implicit form \cite{Beam:1978aa}.
The energy conservation equation is rewritten for $\Delta T_{\rm e}^k\equiv T_{\rm e}^{k+1}-T_{\rm e}^{k}$, where $k$ is the step number.
The discretized equation is written in a linear algebraic form for the $\Delta T_{\rm e}^k$ at the nine stencils (
$\Delta T_{{\rm e,}i+2,j}$, $\Delta T_{{\rm e,}i+1,j}$, 
$\Delta T_{{\rm e,}i,j+2}$, $\Delta T_{{\rm e,}i,j+1}$, 
$\Delta T_{{\rm e,}i,j}$  , $\Delta T_{{\rm e,}i-1,j}$, 
$\Delta T_{{\rm e,}i-2,j}$, $\Delta T_{{\rm e,}i,j-1}$, and
$\Delta T_{{\rm e,}i,j-2}$).
The direct matrix inversion method based on Gaussian elimination is used for the computation of $\Delta T_{\rm e}^k$; then, $T_{\rm e}$ is updated.

\section{Simulation Condition}
An anode-layer-type Hall thruster, UT58 \cite{Bak:2019aa}, is assumed as the simulation target.
The full cylinder of the channel centerline is used and unwrapped to create the x-y two-dimensional calculation domain.
The calculation domain and boundary conditions are shown in Fig. \ref{fig:condition}.
The left-hand-side boundary corresponds to the anode side.
The anode potential $\phi_{\rm a}$ is imposed on the boundary under Dirichlet conditions.
Neutral particles corresponding to the anode mass flow rate were uniformly injected from the anode-side boundary.
The right-hand side boundary is the downstream cathode side.
The cathode-side potential $\phi_{\rm c}$ is also uniformly assumed, and ions and neutral particles flow out from this boundary.
The periodic boundary condition is assumed on the top and bottom boundaries.
The magnetic flux density $B$ is assumed to have only the z- (radial) component.
$B$ has a distribution in the x-direction, while it is assumed to be uniform in the y-direction.
The magnetic field distribution is fixed during the simulation.

The relationship between the calculation domain and thruster walls is shown in Fig. \ref{fig:wall}.
The zero-point in the x-direction is defined at the tip of the hollow anode. The simulation domain is extended to 10 mm inside the hollow anode and 30 mm downstream of the hollow anode tip.
The experimental observations indicated that the main acceleration region exists in the region of 0 mm $<x<$ 30 mm \cite{Bak:2019aa}.

Parameters assumed for the simulation are listed in Table \ref{tab:table1}.
It was reported that the breathing-mode oscillation was small under this operation condition \cite{Bak:2019aa}.
This operation point was selected in this study to focus the present simulation on the analysis of the GDI.
The discharge voltage, anode mass flow rate, and magnetic flux density were set based on the experimental conditions.
The potential and electron temperature at the cathode-side boundary ($x=$ 30 mm) are assumed based on the probe measurement results.

\begin{table}
    \caption{\label{tab:table1}Assumed parameters for the thruster operation and simulation.}
    \begin{ruledtabular}
    \begin{tabular}{lcc}
        Parameter&Symbol&Value\\
        \hline
        Anode potential & $\phi_{\rm a}$ & 150 V\\
        Potential at cathode boundary & $\phi_{\rm c}$ & 40 V\\
        Peak magnetic flux density & $B_{\rm p}$ & 10 mT\\
        Anode mass flow rate & $\dot{m}$ & 2.04 mg/s\\
        Channel wall distance & $w_{\rm ch}$ & 8.0 mm\\
        Hollow anode distance & $w_{\rm HA}$ & 4.0 mm\\
        Electron temperature at cathode boundary & $T_{\rm e,c}$ & 2.7 eV\\
        Time step (PIC) &$\Delta t_{\rm i}$ & 1.0 ns\\
        Time step (electron fluid) & $\Delta t_{\rm e}$ & 0.01 ns\\
    \end{tabular}
    \end{ruledtabular}
\end{table}

Approximately eight million macroparticles are treated in the simulation domain, including ions and neutral particles.
The nominal grid system used in this study was 108 $\times$ 108 with uniform sizing, and the average number of macroparticles per cell was approximately 600.
Each simulation was performed for 200 \textmu s with the time steps for the ion PIC of $\Delta t_{\rm i}=1.0$ ns.
The number of electron subloops is determined based on the stability of the computation.
In the present study, one hundred subloops are used, which means that the time step for the electron fluid is 1/100 of that for the PIC.
One simulation required approximately four days with sequential computation on a single Intel Core i9-9900K 3.6 GHz.

\section{Results and Discussion}

\subsection{Plasma instability}
\label{sec:dist}
\begin{figure*}
    \includegraphics[width=175mm]{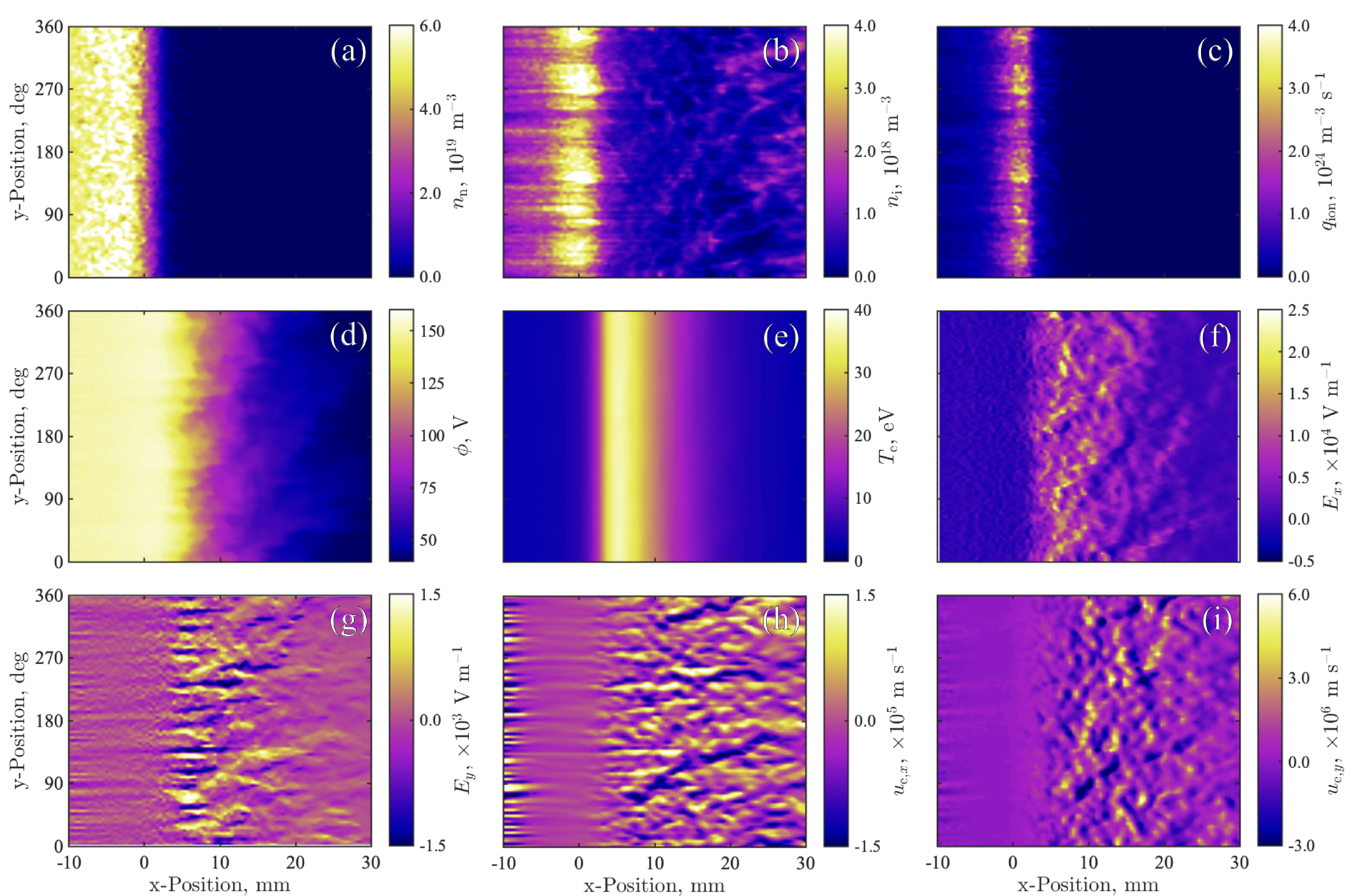}
    \caption{Temporal plasma property distributions simulated by the hybrid model. (a) neutral number density, (b) ion number density, (c) ionization rate, (d) plasma potential, (e) electron temperature, (f) x- (axial) electric field, (g) y- (azimuthal) electric field, (h) x- (axial) electron velocity, and (i) y- (azimuthal) electron velocity. (Multimedia view)}
    \label{fig:dist2d}
\end{figure*}

The simulation was started with the initial condition of azimuthally uniform plasma property distributions.
After approximately 1 \textmu s, plasma instabilities arose, and after approximately 5 \textmu s, the instabilities were saturated.
The temporal two-dimensional plasma property distributions after 200 \textmu s are shown in Fig. \ref{fig:dist2d} (Multimedia view).
The behavior of the plasma instability can be clearly observed in the videos.
These videos show the plasma characteristics during the 25 \textmu s after taking the snapshots in Fig. \ref{fig:dist2d}.
Although the thruster operation condition of small breathing-mode oscillation was selected for the thruster used in this simulation, a breathing-mode with a frequency of 30 kHz was still observed.
The gradual change in the ion number density shown in the video was due to breathing-mode ionization oscillation.
The neutral number density, ionization rate, and electron temperature were relatively stationary.
The distributions of these properties were almost uniform in the y- (azimuthal) direction.
The quiet behaviors of the neutral number density and ionization rate indicated that the instability simulated in this study had little relevance to the ionization oscillation.

\begin{figure}
    \includegraphics[width=80mm]{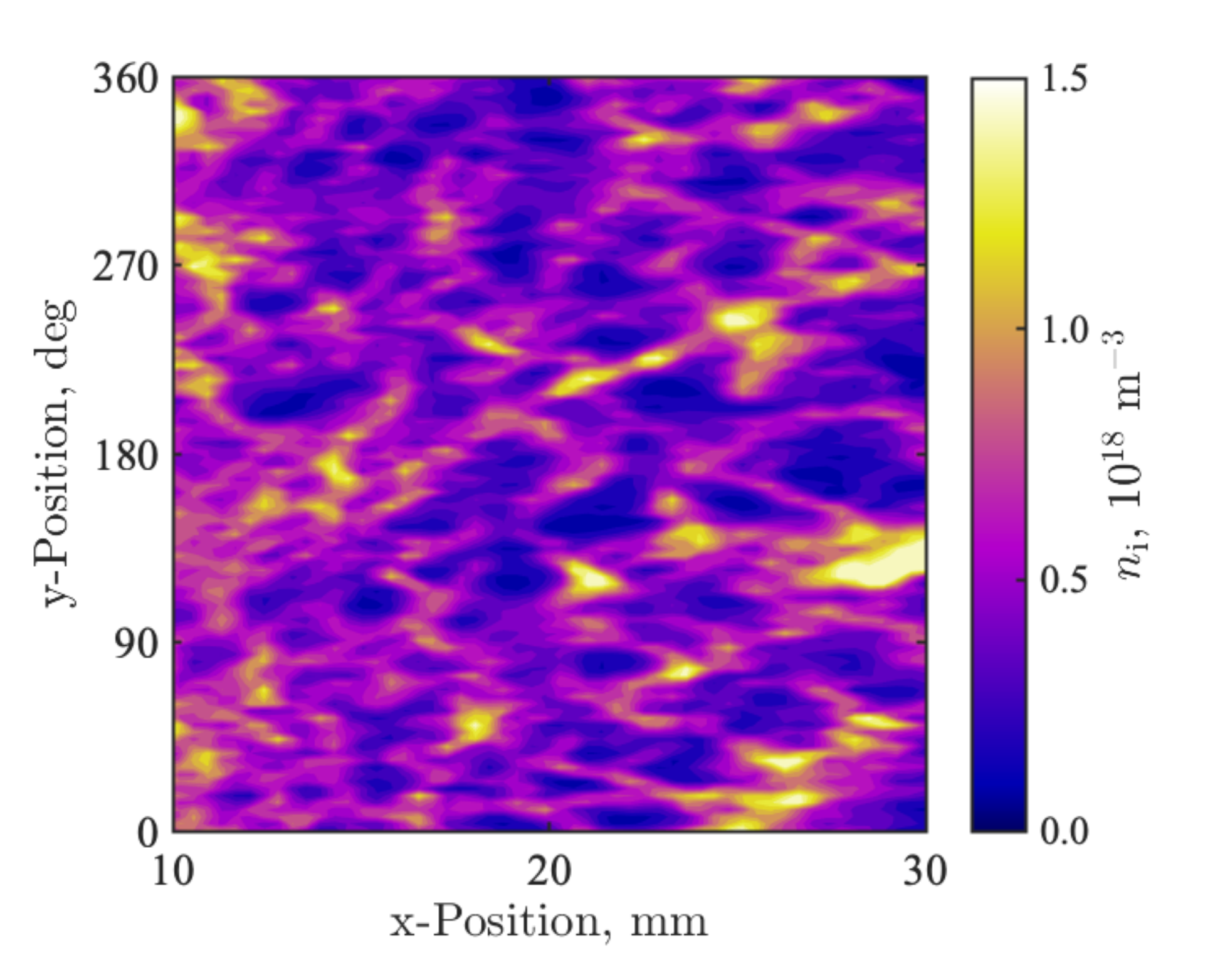}
    \caption{Temporal ion number density distribution in the downstream plume region.}
    \label{fig:ni_plume}
\end{figure}

The ion number density and plasma potential presented strong instabilities.
The behaviors of these instabilities were complicated.
In the region of $-10$ mm $<x\leq$ 2 mm, which was upstream of the acceleration zone, the plasma potential was almost uniform in the y-direction.
Around the acceleration zone of 5 mm $<x\leq$ 15 mm, the plasma potential became unstable with apparently random structures.
The ion flow was also unstable, and the accelerated ion flow in the downstream plume region of 15 mm $<x\leq$ 30 mm was intermittent.
A temporal structure of ion number density in the downstream plume region is visualized in Fig. \ref{fig:ni_plume}.
The intermittent plasma structure was supposed to consist of several waves of different wave numbers, as discussed in Sec. \ref{sec:perturbation}.
The mixed-wave structure essentially moves in the +x-direction according to the accelerated ion velocity.
Because the ionization zone of high $q_{\rm ion}$ did not move in the x-direction during the instability, the intermittent structure was caused by the interplay of plasma density and potential fluctuations, rather than the ionization.

Owing to the unstable plasma potential, the x- and y-electric fields also fluctuated.
The amplitude of the fluctuation in $E_x$ is on the order of 10$^4$ V/m.
In the Hall thruster discharge, the axial electric field in the acceleration zone was typically $\sim 10^4$ V/m.
Therefore, the magnitude of the fluctuation in $E_x$ was on the same order as the stationary part, and the instability had a significant effect on the ion motion.
The amplitude of the fluctuation in $E_y$ is on the order of 10$^3$ V/m, and it is smaller than the amplitude of $E_x$ by one order of magnitude.

The magnetized electron flow is strongly affected by the instability.
Both the x-electron velocity and y-electron velocity are highly oscillatory because of the fluctuating electric fields and electron pressure ($en_{\rm e}T_{\rm e}$).
This instability is induced in the acceleration zone and plume region, where 5 mm $<x\leq$ 30 mm.
In these regions, $\partial B/\partial x<0$ and $\partial n_{\rm e}/\partial x<0$, and the criterion for the induction of GDI \cite{Choueiri:2001ab} is satisfied.
The magnitude of the oscillatory $u_{{\rm e},x}$ is 10$^5$ m/s, and the magnitude for $u_{{\rm e},y}$ is 10$^6$ m/s.
This magnitude relationship is opposite from that for the electric field because the oscillatory part of $u_{{\rm e},x}$ is induced by $E_y$, whereas $u_{{\rm e},y}$ is affected by $E_x$ because of the E$\times$B drift.
An oscillation was observed in the very upstream region of $x\leq -5$ mm.
In this region, $\partial B/\partial x>0$ and $\partial n_{\rm e}/\partial x>0$, and thus, a small GDI is supposed to be induced. However, as shown in Sec. \ref{sec:transport}, the instability in this upstream region has little influence on the cross-field electron transport.

The fluctuation amplitudes of various plasma properties were quantitatively evaluated at x-positions of 0 mm, 8 mm, and 24 mm, as listed in Table \ref{tab:table2}.
These three x-positions correspond to the locations of the peak plasma density, acceleration, and plume zones, respectively.
For instance, the fluctuation amplitude and equilibrium value of the plasma potential were respectively calculated as
\begin{equation}
    \tilde{\phi} = \sqrt{\frac{1}{2\pi}\int^{2\pi}_0 \phi^2 {\rm d}y},
\end{equation}
\begin{equation}
    \bar{\phi}=\frac{1}{2\pi}\int^{2\pi}_0\phi {\rm d}y.
\end{equation}
Furthermore, the oscillatory and stationary quantities were respectively averaged during 30 \textmu s.
In Table \ref{tab:table2}, the fluctuation amplitudes were normalized by the equilibrium quantities, whereas the amplitude of the y- (azimuthal) electric field $\tilde{E}_y$ was not normalized because $\bar{E}_y\approx 0$.
It was found that the oscillatory part of the electron flow velocity was much larger than the stationary part at x-positions of 8 mm and 24 mm.
Thus, the electron flow was significantly affected by the plasma instability, especially in the acceleration and plume regions.

The electron flow exhibits vortex-like structures as visualized in the videos.
This flow field is different from the simulation results of the kinetic ECDI in Hall thrusters, in which a specific mode of azimuthal wave is typically observed  \cite{Adam:2003aa,Lafleur:2017aa}.
The present hybrid particle-fluid simulation is targeted to the fluidic GDI, and the obtained vortex-like flow field reflects the characteristics of the GDI, as discussed in Sec. \ref{sec:perturbation}.
The spatial scale of the vortex-like structures is considered in Sec. \ref{sec:convergence}.

\begin{table}
    \caption{\label{tab:table2}Fluctuation amplitudes of plasma properties normalized by stationary quantities at various x-positions.}
    \begin{ruledtabular}
    \begin{tabular}{cccc}
        $x$ & 0 mm & 8 mm & 24 mm\\
        \   & (peak density) & (acceleration) & (plume) \\
        \hline
        $\tilde{n}_{\rm i}/\bar{n}_{\rm i}$ & 0.24 & 0.21 & 0.53\\
        $\tilde{\phi}/\bar{\phi}$
        & 5.7$\times10^{-3}$& 0.050 & 0.044\\
        $\tilde{T}_{\rm e}/\bar{T}_{\rm e}$ & 7.2$\times10^{-3}$& 5.4$\times10^{-3}$ & 0.016\\
        $\tilde{E}_y$ & 290 V m$^{-1}$& 1600 V m$^{-1}$ & 420 V m$^{-1}$\\
        $\tilde{u}_{{\rm e},x}/|\bar{u}_{{\rm e},x}|$ & 1.2 & 18 & 16\\
    \end{tabular}
    \end{ruledtabular}
\end{table}

\begin{figure}
    \includegraphics[width=80mm]{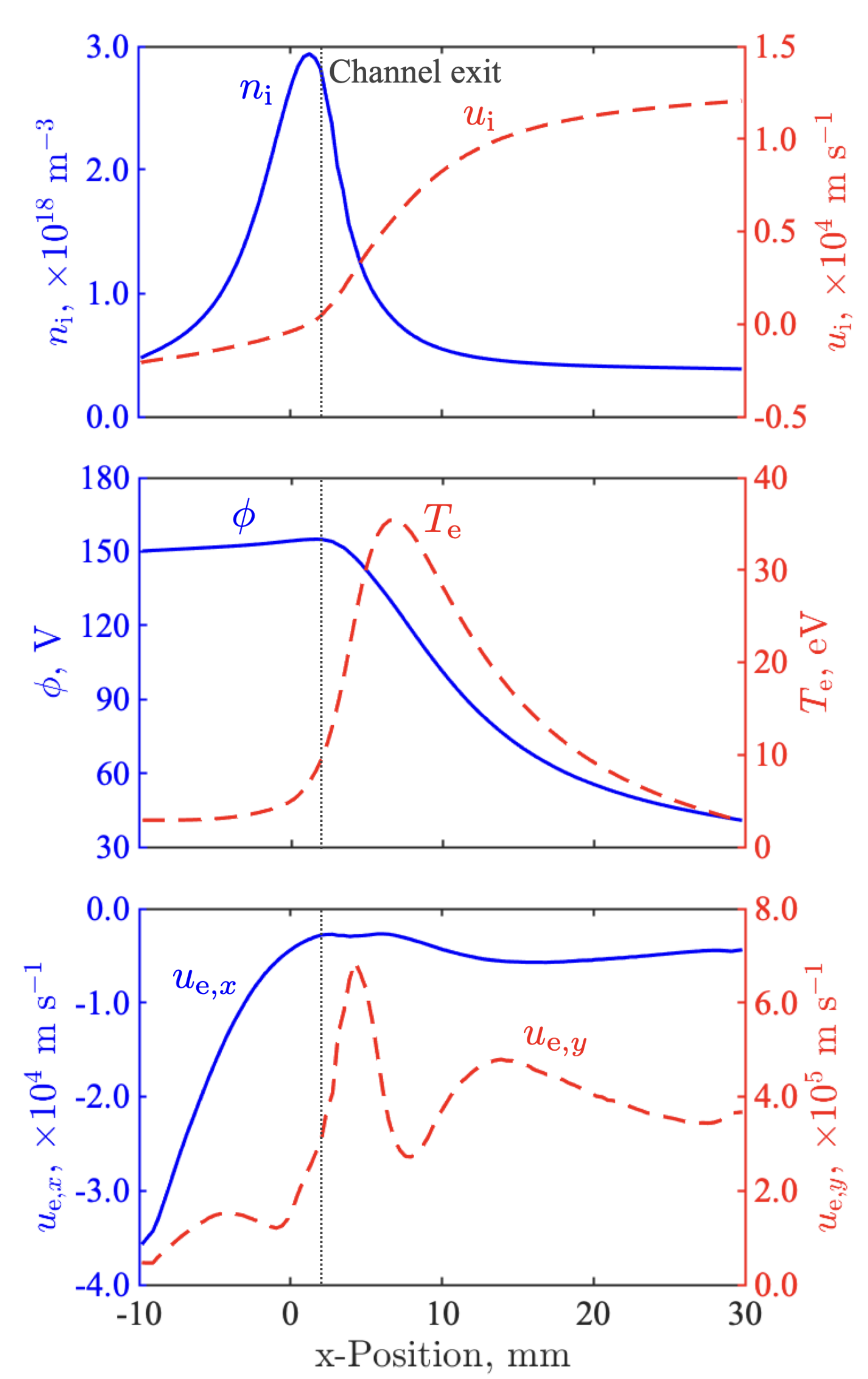}
    \caption{Plasma property distributions in the x-direction obtained by averaging the numerical simulation results in the y-direction.
    [Associated dataset available at http://dx.doi.org/10.5281/zenodo.4695404] (Ref. \onlinecite{KawashimaZenodo2021}).}
    \label{fig:dist1d}
\end{figure}

Azimuthally averaged plasma property distributions in the x- (axial) direction are shown in Fig. \ref{fig:dist1d}.
The peak ion number density is on the order of 10$^{18}$ m$^{-3}$ at $x=1$ mm (middle of the discharge channel).
The ion flow velocity becomes negative in the upstream region of the discharge.
These properties are similar to the typical Hall thruster discharge characteristics \cite{Parra:2006aa}.
The acceleration zone and peak of the electron temperature exist downstream of the channel exit.
These characteristics have been observed in the measurement results of anode-layer-type Hall thrusters \cite{Bak:2019aa,Hamada:2021aa}.

The azimuthally averaged $u_{{\rm e},x}$ is on the order of 10$^4$ m/s, which is smaller than the oscillatory part of $u_{{\rm e},x}$ shown in Fig. \ref{fig:dist2d}(h) by one-order of magnitude.
The azimuthally averaged $u_{{\rm e},x}$ is small in the plume region, where magnetic confinement is relatively strong, whereas the amplitude of the instability becomes large in this region.
Azimuthally averaged $u_{{\rm e},y}$ is on the order of 10$^5$ m/s, and this is smaller than the nonaveraged $u_{{\rm e},y}$ in Fig. \ref{fig:dist2d}(i) by one-order of magnitude.
Azimuthally averaged $u_{{\rm e},y}$ shows a double-peak structure at approximately $x=5$ mm, which indicates that there is a shear in the azimuthal electron motion.
This location is consistent with the region where instability arises.
This type of velocity shear was observed in other numerical simulations in the axial-azimuthal domains \cite{Fernandez2015iepc}, and the shear was considered to be related to the stabilization and saturation of the instability \cite{Burrell:1997aa}.
The formation process and physical meaning of the double-peak structure in the Hall thruster discharge plasma will be considered in a subsequent study.

In summary, plasma instability in the axial-azimuthal domain of a Hall thruster is simulated by a hybrid model with a nonoscillatory electron fluid calculation.
The plasma instability exhibits a complicated plasma structure, which is considered to consist of several waves of different wave numbers.
Perturbation analyses are useful for investigating the characteristics of the plasma instability, such as the growth rate.
Therefore, we performed linear perturbation analyses of the simulated plasma properties in Sec. \ref{sec:perturbation}.
The simulation results were compared with perturbation theories to validate the present numerical model for the GDI.

\subsection{Electron transport enhancement}
\label{sec:transport}
The effects of the simulated plasma instability on the axial cross-field electron transport are discussed in this section.
Because the kinetic ECDI is not included in the hybrid model used in this study, the electron transport enhancement induced solely by the GDI was analyzed. 
In the drift-diffusion equation for the electron fluid, the electron current density in the x- (axial) direction $j_{{\rm e},x}$ is expressed as
\begin{align}
    j_{{\rm e},x}=&j_{{\rm e},x,{\rm diffusion}}+j_{{\rm e},x,{\rm drift}}\nonumber \\
    =&\mu_\perp\left(-en_{\rm e}\frac{\partial \phi}{\partial x}+\frac{\partial}{\partial x}\left(en_{\rm e}T_{\rm e}\right)\right)\nonumber \\
    +&\mu_\perp\Omega_{\rm e}\left(-en_{\rm e}\frac{\partial \phi}{\partial y}+\frac{\partial}{\partial y}\left(en_{\rm e}T_{\rm e}\right)\right),
    \label{eq:jex1}
\end{align}
where $j_{{\rm e},x,{\rm diffusion}}$ is induced from the x- (axial) gradients, and it corresponds to classical diffusion.
$j_{{\rm e},x,{\rm drift}}$ is from the y- (azimuthal) gradients, and it corresponds to the E$\times$B and diamagnetic drifts.
By taking the azimuthal average of Eq. (\ref{eq:jex1}), the net electron current density can be defined as
\begin{align}
    \bar{j}_{{\rm e},x}=&\bar{j}_{{\rm e},x,{\rm diffusion}}+\bar{j}_{{\rm e},x,{\rm drift}}.
    \label{eq:jex2}
\end{align}
If one considers the net electron current flowing in the x-direction, the effect of diamagnetic drift (pressure gradient in the y-direction) is canceled when it is integrated in the y- (azimuthal) direction.
Thus, only the E$\times$B drift (potential gradient in the y-direction) contributes to $\bar{j}_{{\rm e},x,{\rm drift}}$.
Furthermore, the effective electron mobility is considered based on the simulation results. In axisymmetric simulations of E$\times$B plasma devices, one cannot handle the gradients in the azimuthal direction. In this case, the cross-field electron mobility is artificially modified to include the effects of electron transport enhancement induced by E$\times$B drift, which is known as the effective electron mobility $\mu_{\perp,{\rm eff}}$. It would be valuable to extract $\mu_{\perp,{\rm eff}}$ based on the present simulation results. In this study, $\mu_{\perp,{\rm eff}}$ is calculated using the net electron current and the y-averaged plasma quantities as
\begin{align}
    \bar{j}_{{\rm e},x}=\mu_{\perp,{\rm eff}}\left(-e\bar{n}_{\rm e}\frac{\partial \bar{\phi}}{\partial x}+\frac{\partial}{\partial x}\left(e\bar{n}_{\rm e}\bar{T}_{\rm e}\right)\right).
    \label{eq:jex2}
\end{align}
Here, because only the gradients in the x-direction are included on the right-hand side, the contribution from $\bar{j}_{{\rm e},x,{\rm drift}}$ is represented by $\mu_{\perp,{\rm eff}}$.
The effective electron Hall parameter $\Omega_{\rm eff}$ is defined by the relationship between the x- and y-electron current densities as
\begin{equation}
    \Omega_{\rm eff}\equiv\left|\bar{j}_{{\rm e},y}/\bar{j}_{{\rm e},x}\right|.
    \label{eq:jex3}
\end{equation}

\begin{figure}
    \includegraphics[width=75mm]{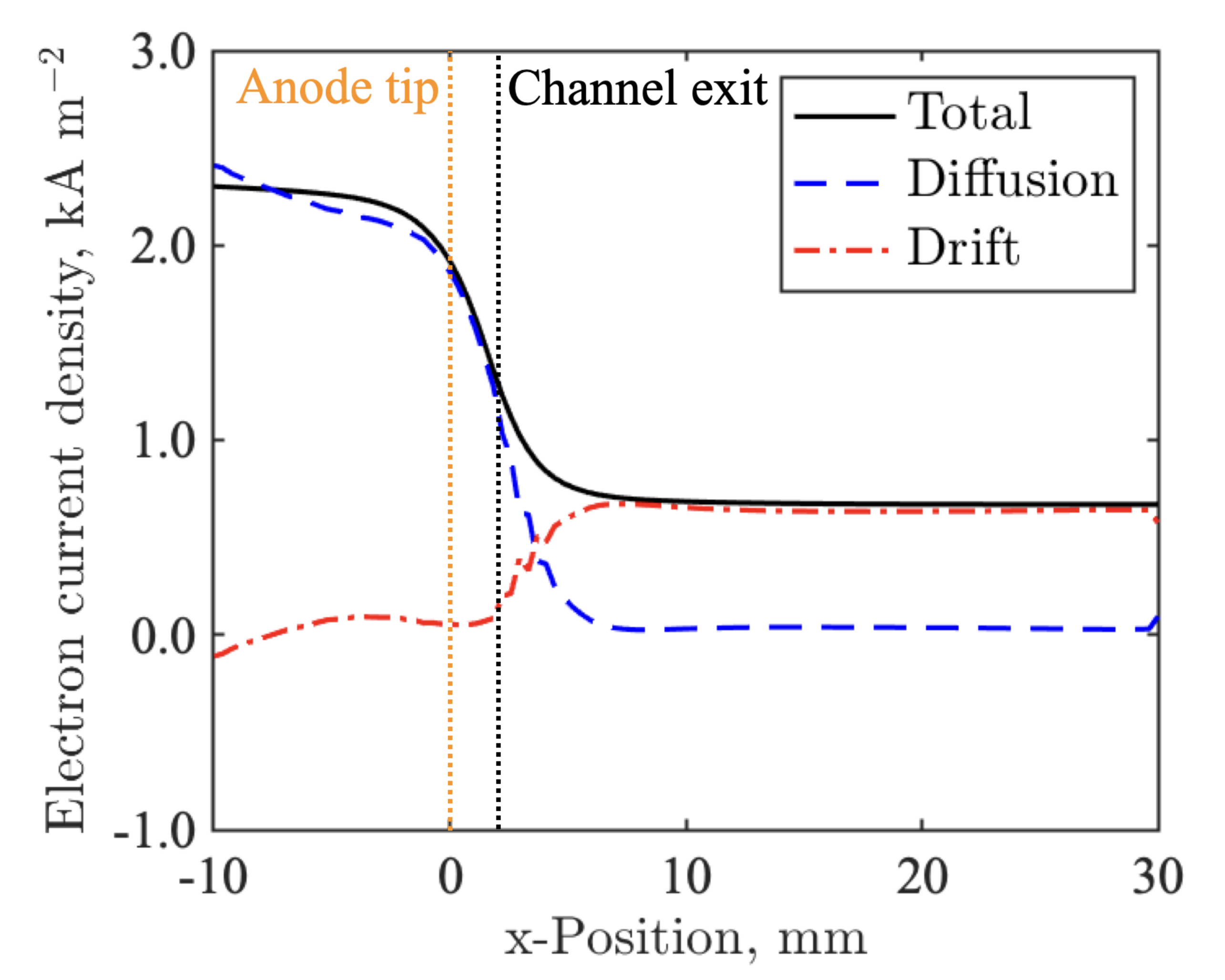}
    \caption{Axial electron current density distribution. Contributions from the diffusion and drift effects are also plotted.
    [Associated dataset available at http://dx.doi.org/10.5281/zenodo.4695404] (Ref. \onlinecite{KawashimaZenodo2021}).}
    \label{fig:jex1d}
\end{figure}

The x-distribution of the net electron current density $\bar{j}_{{\rm e},x}$ is shown in Fig. \ref{fig:jex1d}.
The $\bar{j}_{{\rm e},x}$ at the right-hand side boundary corresponds to the electrons coming from the cathode to the calculation domain.
Here, the net electron flow is oriented toward the anode.
After passing through the main ionization zone around the channel exit, $\bar{j}_{{\rm e},x}$ increases.
Finally, electrons flow out to the anode at the left-hand side boundary, which is the main factor of the discharge current.
Electron flux conservation is strictly calculated in the present simulation.
The balance among the cathode-side electron current, anode-side electron current, and the generated electron amount per unit time, is strictly satisfied with an error of less than 10$^{-9}$ A, while the discharge current is around 2.5 A.

In Fig. \ref{fig:jex1d}, contributions from the diffusion effect ($\bar{j}_{{\rm e},x,{\rm diffusion}}$) and drift effect ($\bar{j}_{{\rm e},x,{\rm drift}}$) are also plotted.
Inside the hollow anode, the diffusion effect is predominant in the electron transport because the neutral atom density is large inside the hollow anode, and the classical electron diffusion yields sufficient electron transport toward the anode.
At around the channel exit, $\bar{j}_{{\rm e},x,{\rm drift}}$ starts to increase downstream, whereas $\bar{j}_{{\rm e},x,{\rm diffusion}}$ decreases.
In the acceleration and plume regions where 5 mm $<x$, the drift effect overwhelms the diffusion effect, and the $\bar{j}_{{\rm e},x,{\rm drift}}$ is the predominant factor in the cross-field electron current.
The drift effect is caused by the y-electric field that is generated solely by plasma instability.
Further, it is shown that a significant cross-field electron flux is induced by plasma instability in the acceleration and plume regions. 

\begin{figure}
    \includegraphics[width=80mm]{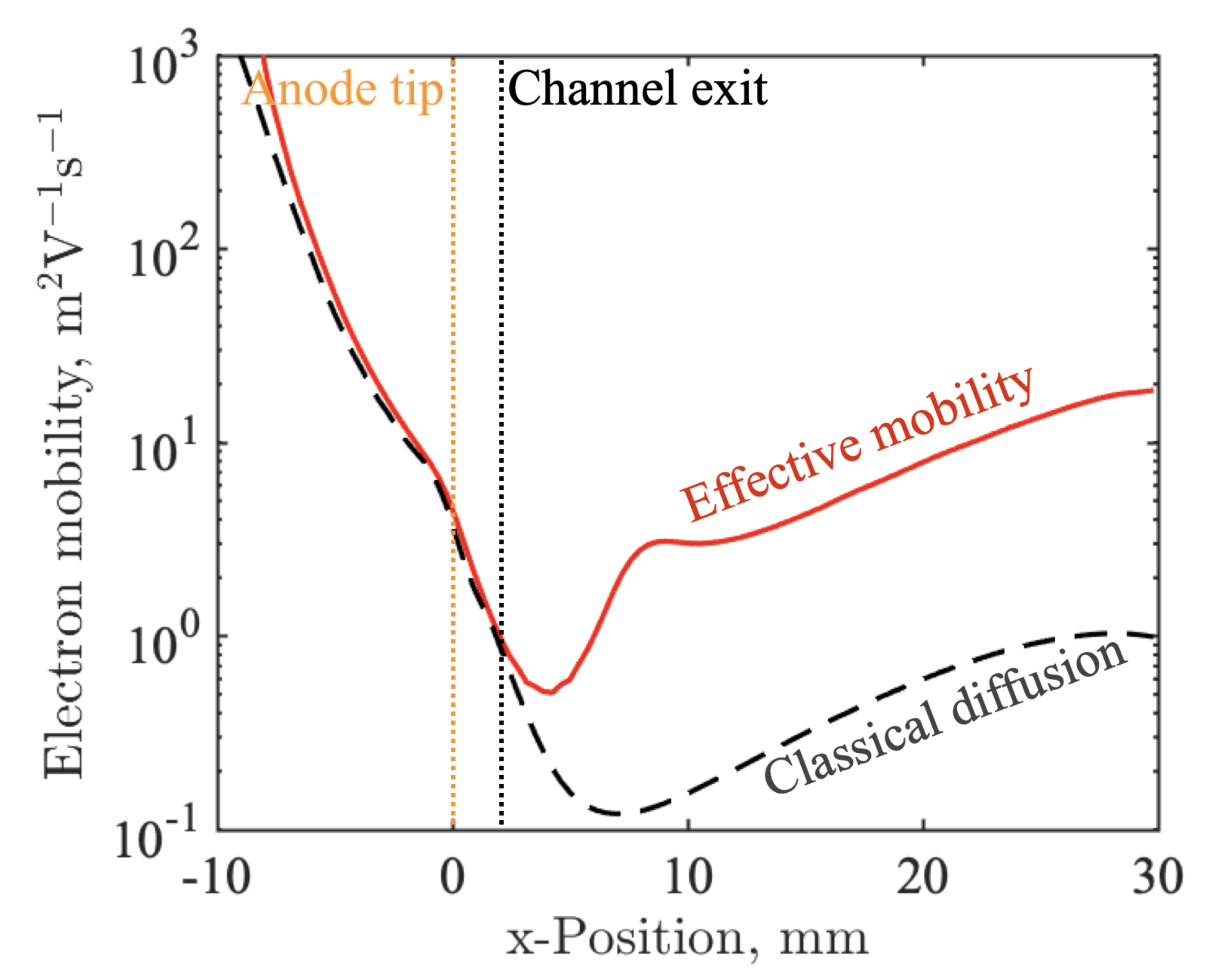}
    \caption{\label{fig:mueff} Effective electron mobility distribution. The electron mobility from classical diffusion is plotted for comparison.
    [Associated dataset available at http://dx.doi.org/10.5281/zenodo.4695404] (Ref. \onlinecite{KawashimaZenodo2021}).}
\end{figure}

The effective electron mobility is shown in Fig. \ref{fig:mueff} as a function of the x-position.
For reference, the azimuthally averaged classical electron mobility obtained from the simulation is also plotted.
Similar to the electron current density, the cross-field electron transport inside the hollow anode is well represented by the classical electron mobility.
The classical electron mobility continues to decrease from the channel exit towards downstream because of the strong magnetic flux density and depleted neutral atom density.
The effective electron mobility becomes greater than the classical mobility in accordance with plasma instability induction.
In the downstream region, where the drift effect overwhelms the diffusion effect, the effective electron mobility is greater than the classical mobility by one order of magnitude.
The trend of the effective electron mobility obtained in this study is qualitatively similar to the empirical data \cite{Lafleur:2016ab}.

\begin{figure}
    \includegraphics[width=80mm]{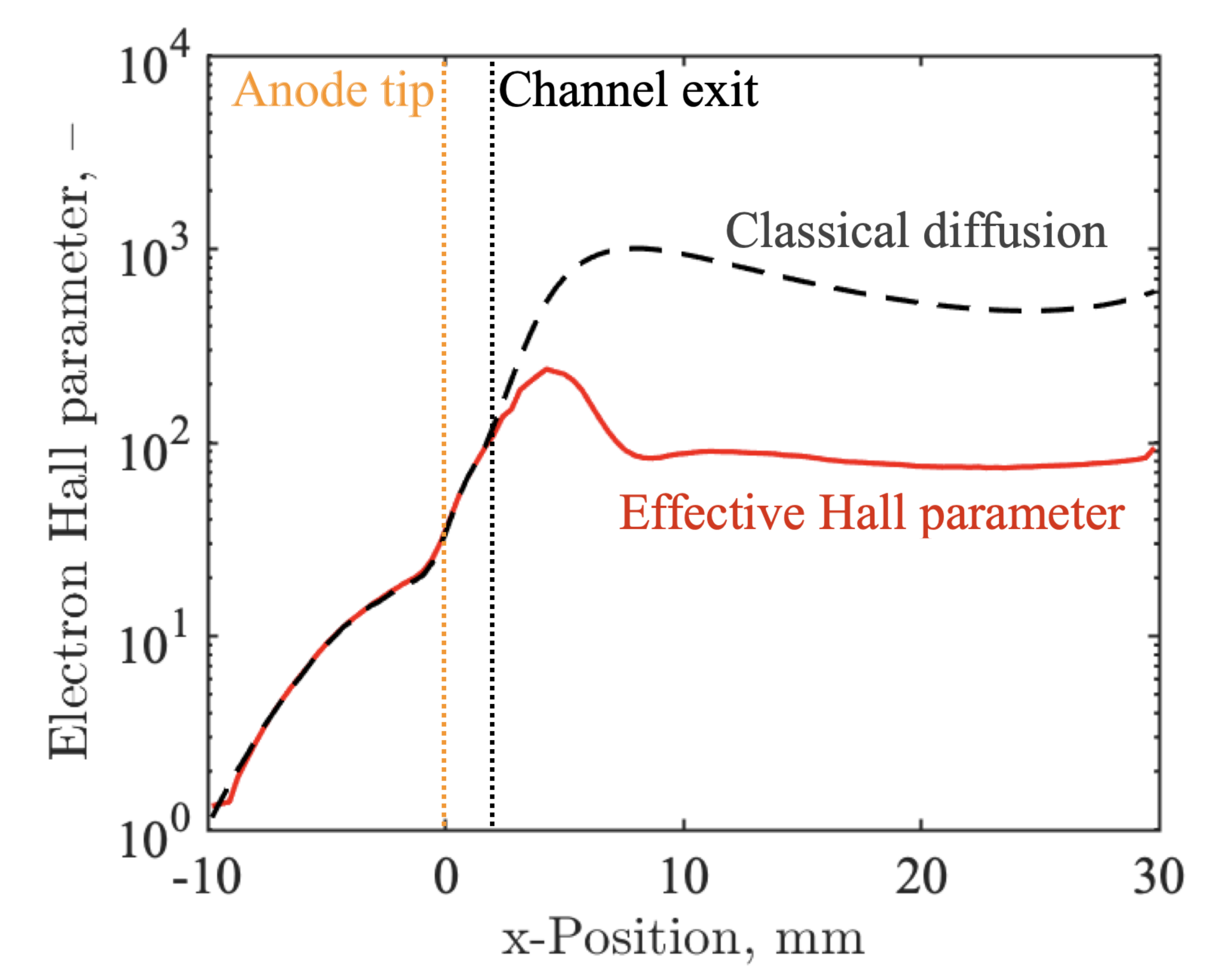}
    \caption{\label{fig:Ome_eff} Effective electron Hall parameter distribution. The Hall parameter from classical diffusion is plotted for comparison.
    [Associated dataset available at http://dx.doi.org/10.5281/zenodo.4695404] (Ref. \onlinecite{KawashimaZenodo2021}).}
\end{figure}

The effective Hall parameters are shown in Fig. \ref{fig:Ome_eff} as a function of the x-position.
Here, the azimuthally averaged classical Hall parameter, expressed as Eq. (\ref{eq:Omega}), is also plotted.
The effective Hall parameter is almost the same as that of the classical diffusion, in the region inside the hollow anode.
However, the effective Hall parameter is smaller than the classical Hall parameter in the acceleration zone and plume regions downstream of the channel exit.
The reduced Hall parameters indicate that the magnetic confinement of the electrons becomes weaker than the theoretical expectation of the classical diffusion theory because of the effects of plasma instability.
The simulated effective Hall parameter is $100<\Omega_{\rm eff}<300$ in the acceleration zone, and the Hall parameter in this range is also observed in the measurement results \cite{Hofer:2006aa}.

In summary, the plasma instability simulated by the hybrid particle-fluid model enhances the cross-field electron transport in the axial direction, especially in the downstream plume region.
In addition, the plasma instability reduces the effective Hall parameter, which indicates that the magnetic confinement of electrons is weakened.
To simulate the cross-field electron transport more accurately and reproduce the experimental measurement, radial effects including diffusive ion flow in the plume region need to be considered.
In the present study, the effects of the radial expansion of the plasma are neglected to focus the model on axial-azimuthal instability.
Further detailed quantitative validation of the cross-field electron transport should be conducted in a future study, such as that in Ref. \onlinecite{KawashimaIEPC2019}, and it should include the radial effects.

\begin{figure*}
    \includegraphics[width=175mm]{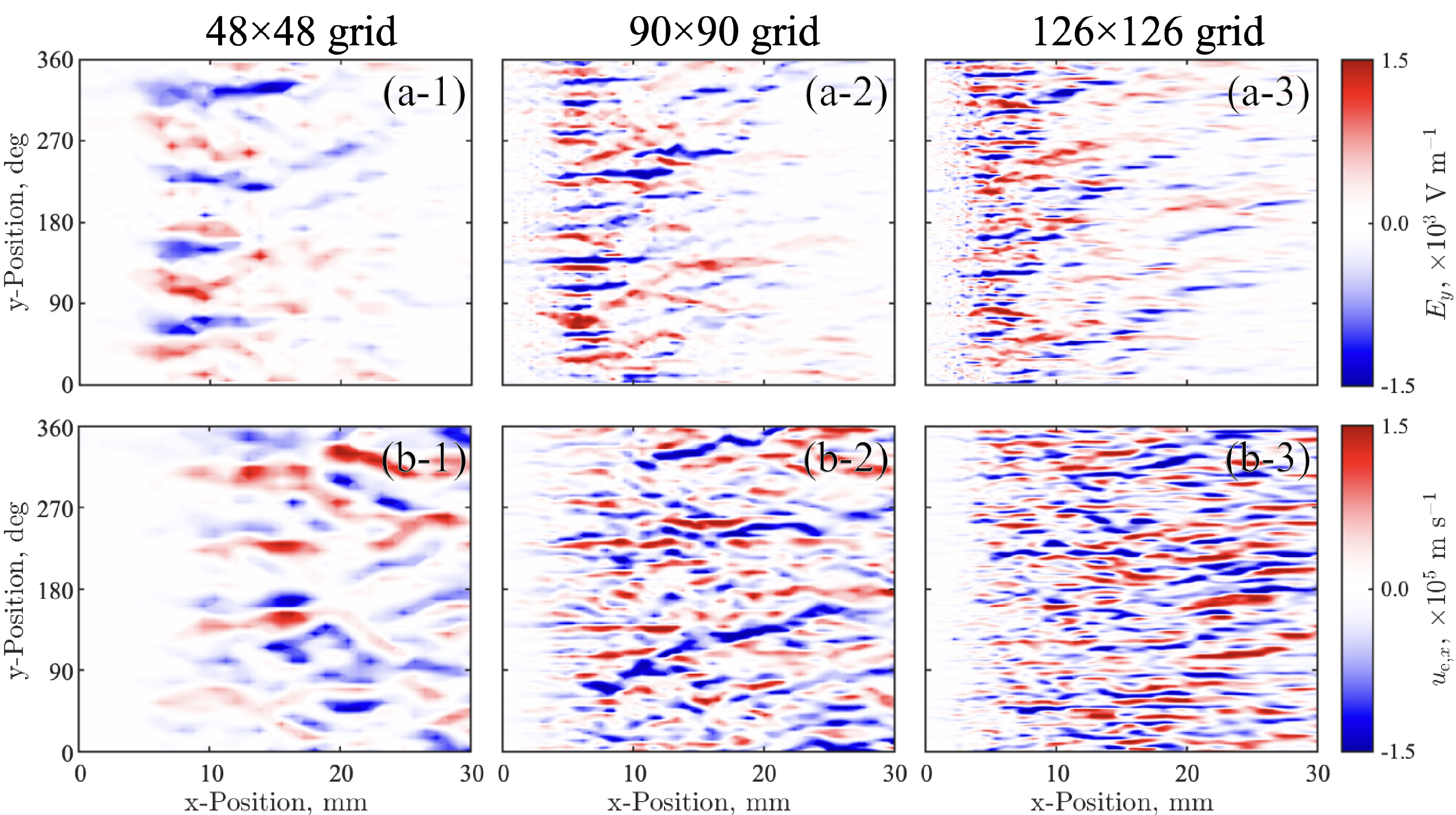}
    \caption{Temporal distributions of (a) y- (azimuthal) electric field and (b) x- (axial) electron velocity simulated with the grid systems of 48$\times$48, 90$\times$90, and 126$\times$126. Colorbars are common among the results of different grid systems.}
    \label{fig:conv_dist}
\end{figure*}

\subsection{Grid convergence}
\label{sec:convergence}
The computational accuracy on the electron transport property is considered.
The governing equation for solving the potential in Eq. (\ref{eq:pot1}) is the so-called anisotropic diffusion equation, and it is known that a calculation for this type of equation tends to include a large discretization error \cite{Kawashima201559,Chamarthi:2018aa}.
In the case of the Z-$\theta$ simulation of Hall thrusters, the magnitude of non-diagonal elements in the mobility tensor is not so large ($\Omega_{\rm e}\sim 10^3$).
However, the discretization error may result in the inaccurate analysis of the cross-field electron transport.
In addition, numerical diffusion is introduced in the potential solver as described in Sec. \ref{sec:pot}.
Thus, it is necessary to evaluate the grid convergence in terms of the instability and electron transport properties.

The 2D distributions of the y-(azimuthal) electric field and x-(axial) electron velocities simulated with grid systems of 48$\times$48, 90$\times$90, and 126$\times$126 are shown in Fig. \ref{fig:conv_dist}.
The spatial size of the oscillation is smaller in the results of finer grid systems.
The fine grid system of 126$\times$126 resolves smaller vortex structures in the electron flow compared to the results from the other grid systems.
Thus, the size of the vortices in the GDI shows a dependence on the grid, and a grid convergence is not obtained.
There is a possibility that the model for providing the minimum scale of the vortices is not included in the present model.
Because plasma instability is generated in the magnetized electron flow, the scale of the vortex structures may be related to the electron Larmor radius.
However, the finite Larmor radius effect is not included in the electron fluid model.
In turbulent flow simulations, the minimum scale of the vortex is called the Kolmogorov scale, which is dependent on the viscosity of the flow.
The present electron fluid model is based on the simplified drift-diffusion equation, and the viscosity tensor is not included.
In Ref. \onlinecite{Smolyakov:2016aa}, the authors reported that the finite electron Larmor radius effects on the GDI can be considered via electron gyroviscosity.
Therefore, if the gyroviscosity tensor is included in the electron fluid model, the Kolmogorov scale of GDI may appear in the simulation of the hybrid model.

Another finding in Fig. \ref{fig:conv_dist} is that the amplitude of the oscillatory $E_y$ and $u_{{\rm e},x}$ do not significantly differ among the different grid-resolution cases.
The fact that the amplitude of $E_y$ does not change implies that in the coarse grid of 48$\times$48, the plasma potential $\phi$ has a relatively large oscillation amplitude with a large spatial scale, whereas the results of fine grid systems show small $\phi$ oscillations with small spatial scales.
Because the amplitude of $u_{{\rm e},x}$ depends on $E_y$, $u_{{\rm e},x}$ also shows a small dependence on the grid resolution.

\begin{figure}
    \includegraphics[width=80mm]{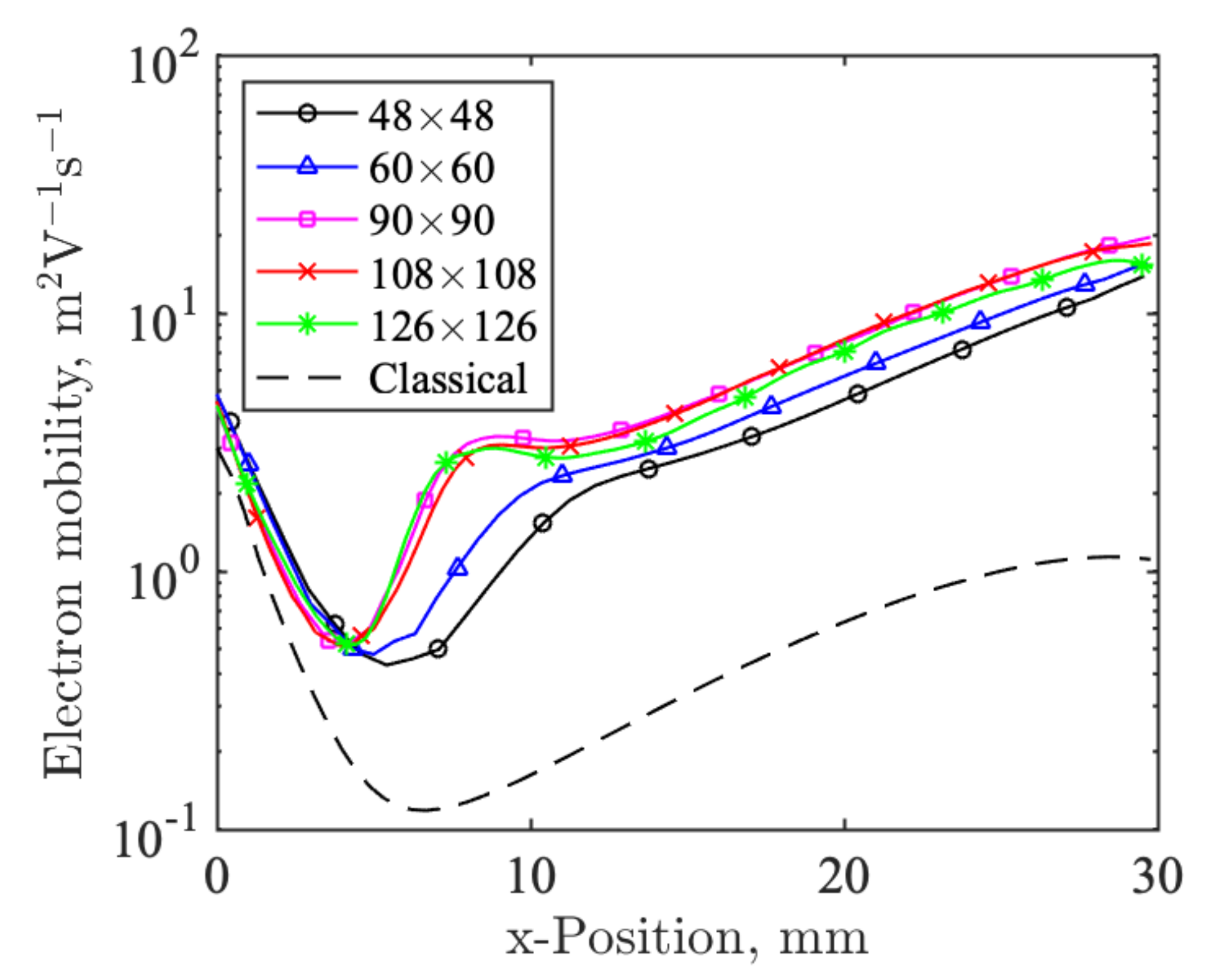}
    \caption{Effective electron mobility distributions simulated with various computational grids.
     The electron mobility from the classical diffusion theory is also plotted for comparison.}
    \label{fig:conv_mob}
\end{figure}

The distributions of effective electron mobility $\mu_{\perp,{\rm eff}}$ in Eq. (\ref{eq:jex2}) are plotted in Fig. \ref{fig:conv_mob} for several grid systems.
In all the results of the different grid systems, the cross-field electron transport is significantly enhanced from the classical electron mobility.
As shown in the results of the coarse grid systems of 48$\times$48 and 60$\times$60, the $\mu_{\perp,{\rm eff}}$ is underestimated in all calculation regions, compared with the results of the finer grid systems.
This is because the coarse grid systems cannot resolve small-scale instabilities, and thus, the effect of electron transport enhancement is partially obscured.
Simulations with fine grid systems of 90$\times$90, 108$\times$108, and 126$\times$126 yield consistent distributions of $\mu_{\perp,{\rm eff}}$.
As shown in Fig. \ref{fig:conv_dist}, the scale of the vortices that can be resolved in these grid systems is slightly different.
However, the very small vortices whose sizes are the sub-grid scale of 90$\times$90 have little effect on the electron transport enhancement.
From this result, one can consider that resolving the minimum scale of the instability may not be necessary for analyzing effective electron transport; further, the simulation for large eddies can be used for electron transport analysis.

The grid convergence of $\mu_{\perp,{\rm eff}}$ observed in Fig. \ref{fig:conv_mob} indicates that the effect of the numerical diffusion is not significant in the cases of fine grid systems, and the simulated electron transport property is based on the physics of the GDI.
The fact that the enhanced electron transport property is obtained with a reasonable number of grids in the whole circumference domain indicates the possibility of further use of the Z-$\theta$ hybrid model.
First, the obtained electron transport property is used as an anomalous electron transport model in Z-R axisymmetric simulations.
One can also consider a coupled simulation of the Z-$\theta$ and Z-R models for high-fidelity simulations of the Hall thruster discharge \cite{KawashimaIEPC2019}, with the small computational load of the hybrid model.
Further, the entire circumference simulation may enable in-depth analyses of long-wavelength rotating spokes \cite{Kawashima:2018ab,Sekerak:2016aa} and thruster operations with artificial inhomogeneities \cite{Bak:2019aa,Bak:2020aa}.

\subsection{Comparison with perturbation theories}
\label{sec:perturbation}
The simulation results were compared with perturbation theories to investigate the relationship between the simulated plasma instability and the theory of GDI.
Linear perturbation analyses were performed based on the stationary plasma profiles obtained from the simulation.
The growth of the simulated plasma instability was compared to that obtained by linear analyses.
Procedures for applying linear analysis are as follows.
(1) The axial profile of the time-averaged plasma distributions is assumed to be the zeroth-order stationary profile.
(2) The azimuthal wave mode number is obtained by applying a fast Fourier transform (FFT) analysis to the plasma instability in the numerical simulation.
(3) The zeroth-order stationary profile is inserted into the dispersion relations based on the GDI theory, which are evaluated locally.
(4) The local growth rate of the dispersion relation was assessed for the assumed azimuthal wave mode number.

In this study, linear and local perturbation models proposed by Esipchuk and Tilinin \cite{Esipchuk:1976} and Frias et al. \cite{FriasPoP2012} were used for perturbation analyses.
The dispersion relations in these perturbation models were summarized in Ref. \onlinecite{Kawashima:2018ab}.
In brief, Esipchuk's model is based on a model of quasineutral, two-fluid, and collisionless plasma without electron inertia.
This model reflects the fundamental characteristics of the GDI, driven by a combination of magnetic flux and plasma density gradients.
Frias's model is also a GDI model that assumes quasineutral, two-fluid, and inertialess electrons, which considers the compressibility of electrons \cite{FriasPoP2012}.
The dispersion relations in these models analyze perturbative solutions that grow on the stationary profile of the plasma properties for an assumed wave mode number.
The x- (axial) distributions of the time-averaged simulation results shown in Fig. \ref{fig:dist1d} were assumed as the stationary zeroth-order profile.

\begin{figure}
    \includegraphics[width=80mm]{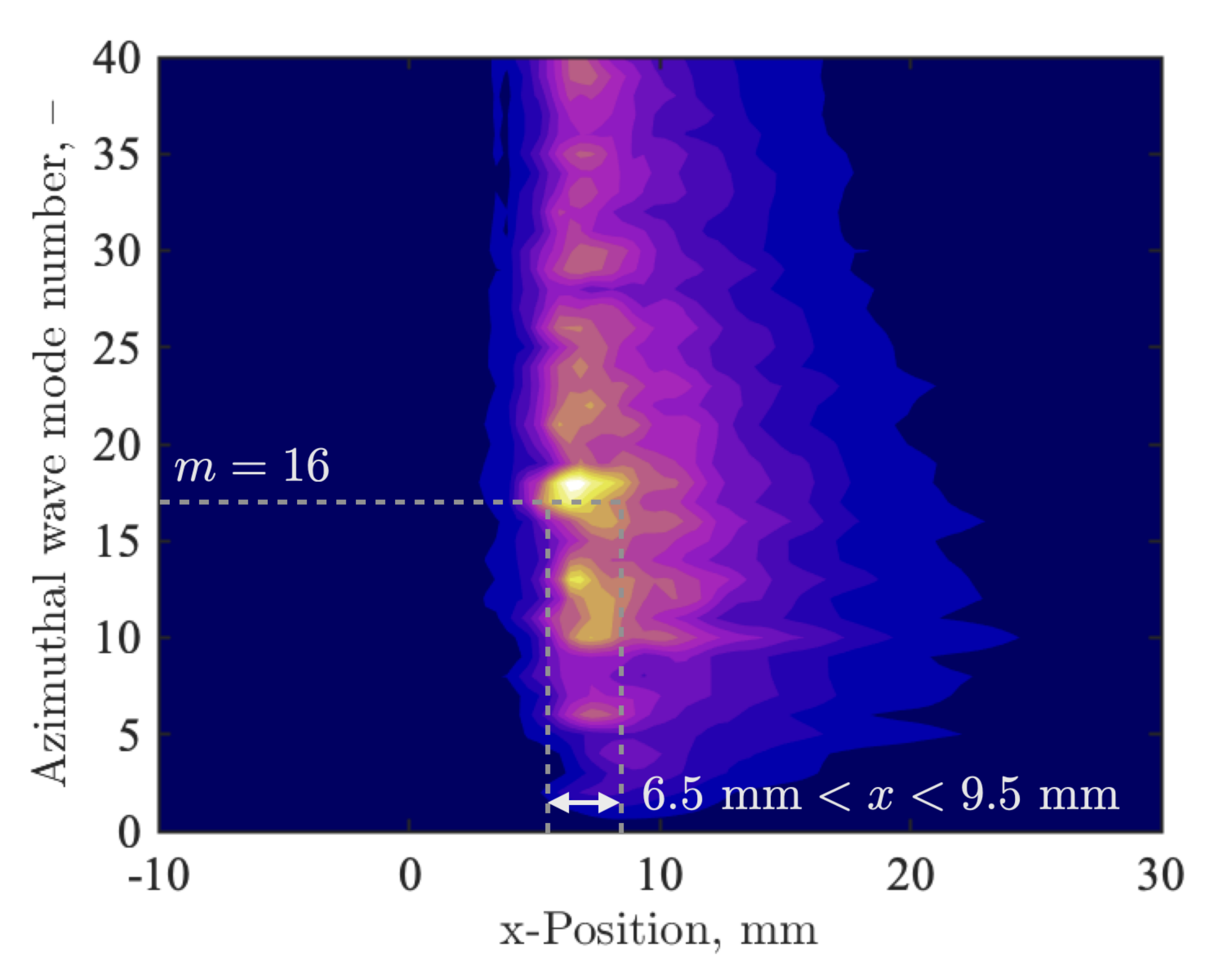}
    \caption{Spectral densities of the simulated plasma instability for various azimuthal wave modes and x-positions. The FFT is applied to the y-distributions of $E_y$ and averaged over 5 \textmu s.}
    \label{fig:fft_Ey}
\end{figure}

In addition, the azimuthal wave mode number $m$ was obtained from the simulation results.
$m$ corresponds to the number of waves in the full circumference of the discharge channel.
An FFT was applied to the y- (azimuthal) direction of $E_y$ to estimate the wave mode number.
The spectral density was obtained for each temporal distribution of $E_y$, and then time-averaged quantities were obtained for the dataset during 5 \textmu s.
The spectral densities of various azimuthal wave modes and x-position are shown in Fig. \ref{fig:fft_Ey}, with an arbitrary unit.
A high spectral density was observed in the region of 6.5 mm $<x<$ 9.5 mm, which corresponds to the acceleration zone.
The existence of various mode numbers ranging 5 $<m<$ 40 was confirmed.
As discussed in Sec. \ref{sec:dist}, the simulation showed a complicated plasma structure. 
The FFT results indicate that the plasma flow consists of several waves with various wave numbers.
However, the peak was detected at $m=16$ in the FFT result, and this wave mode was considered to be the primary component of the instability.
Hence, the azimuthal wave mode number was assumed to be $m=16$ for the perturbation analyses.
This wave mode corresponds to a wave number in the y-direction of $k_y=$ 550 rad/m.
Marusov et al. discussed the general features of the GDI in the Hall thruster discharge and argued that the most unstable mode in the plume region should have a wave number of $k_y\simeq 70-2610$ rad/m \cite{Marusov}.
Thus, the characteristics of the primary wave mode in the numerical simulation were consistent with the proposed general feature of the GDI.

\begin{figure}
    \includegraphics[width=80mm]{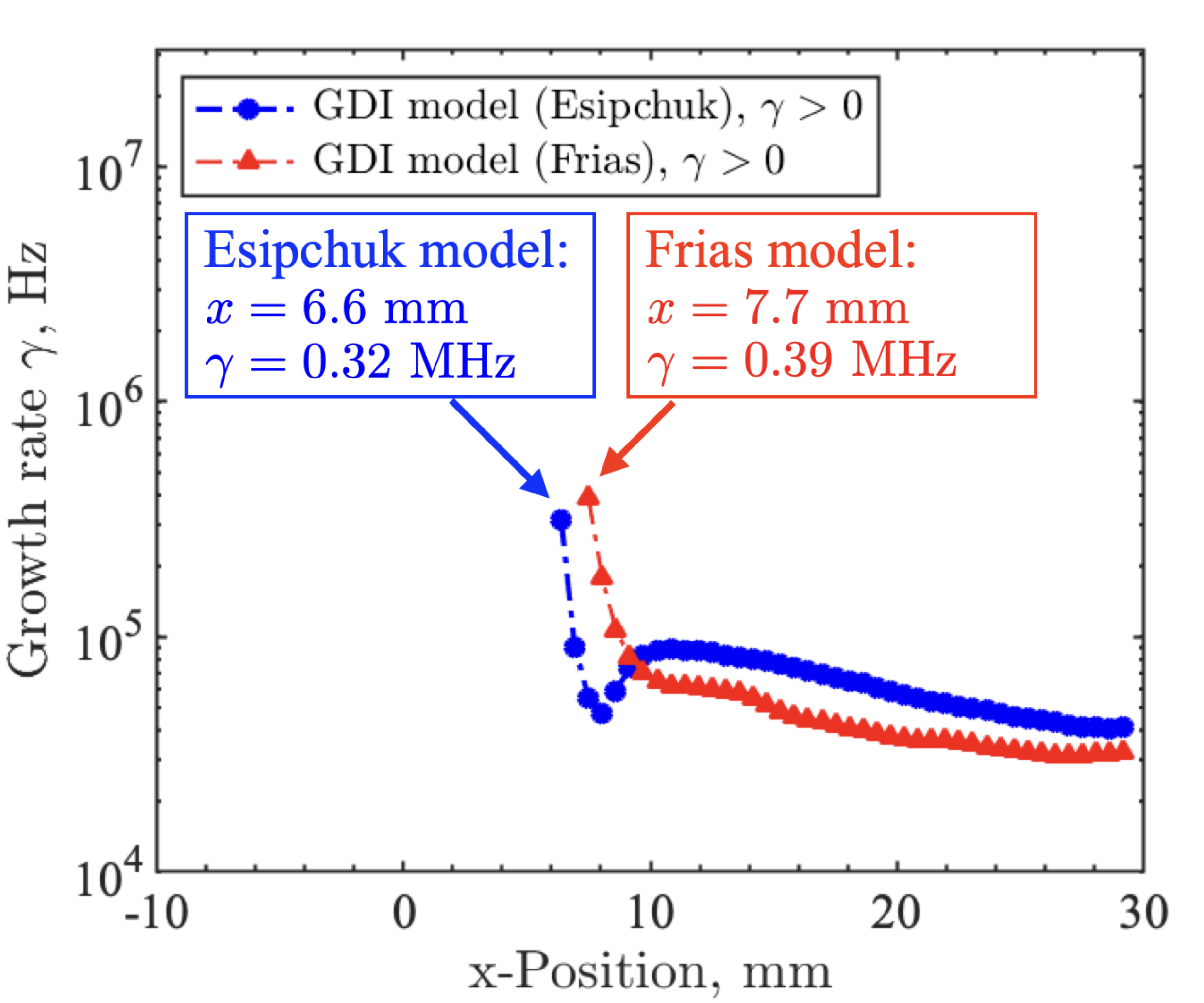}
    \caption{Growth rate distributions obtained from the linear GDI models. Only positive growth rate points are plotted.}
    \label{fig:growth}
\end{figure}

Figure \ref{fig:growth} shows the growth rate distributions obtained by the linear perturbation models.
Note that only the points with positive growth rates are plotted in this figure.
The perturbation models indicate that the GDI rises in the acceleration and plume regions where $x>$ 6.6 mm.
The peak growth rates were 0.32 MHz at $x=$ 6.6 mm and 0.39 MHz at $x=$ 7.7 mm in the results of Esipchuk and Frias models, respectively.
These points are the origin of the GDI in the perturbation models.
Furthermore, these points are included in the high-spectral-density region in Fig. \ref{fig:fft_Ey}.
Therefore, it was concluded that the GDI observed in the numerical simulation agreed with the perturbation theory in terms of the source location of the instability.

The growth rates were compared between the numerical simulation and perturbation models.
To analyze the growth of the instability in the numerical simulation, the simulation was restarted with an initial condition of a uniform plasma property in the y-direction, with the x-distributions of the stationary plasma profile in Fig. \ref{fig:dist1d}.
The amplitude of the simulated oscillation was evaluated by the $\tilde{E}_y$.
$\tilde{E}_y$ was calculated in the same manner as the one in Table \ref{tab:table2}.
Then, $\tilde{E}_y$ was averaged in the region of 6.5 mm $<x<$ 9.5 mm, where the spectral density was high, as shown in Fig. \ref{fig:fft_Ey}.

\begin{figure}
    \includegraphics[width=80mm]{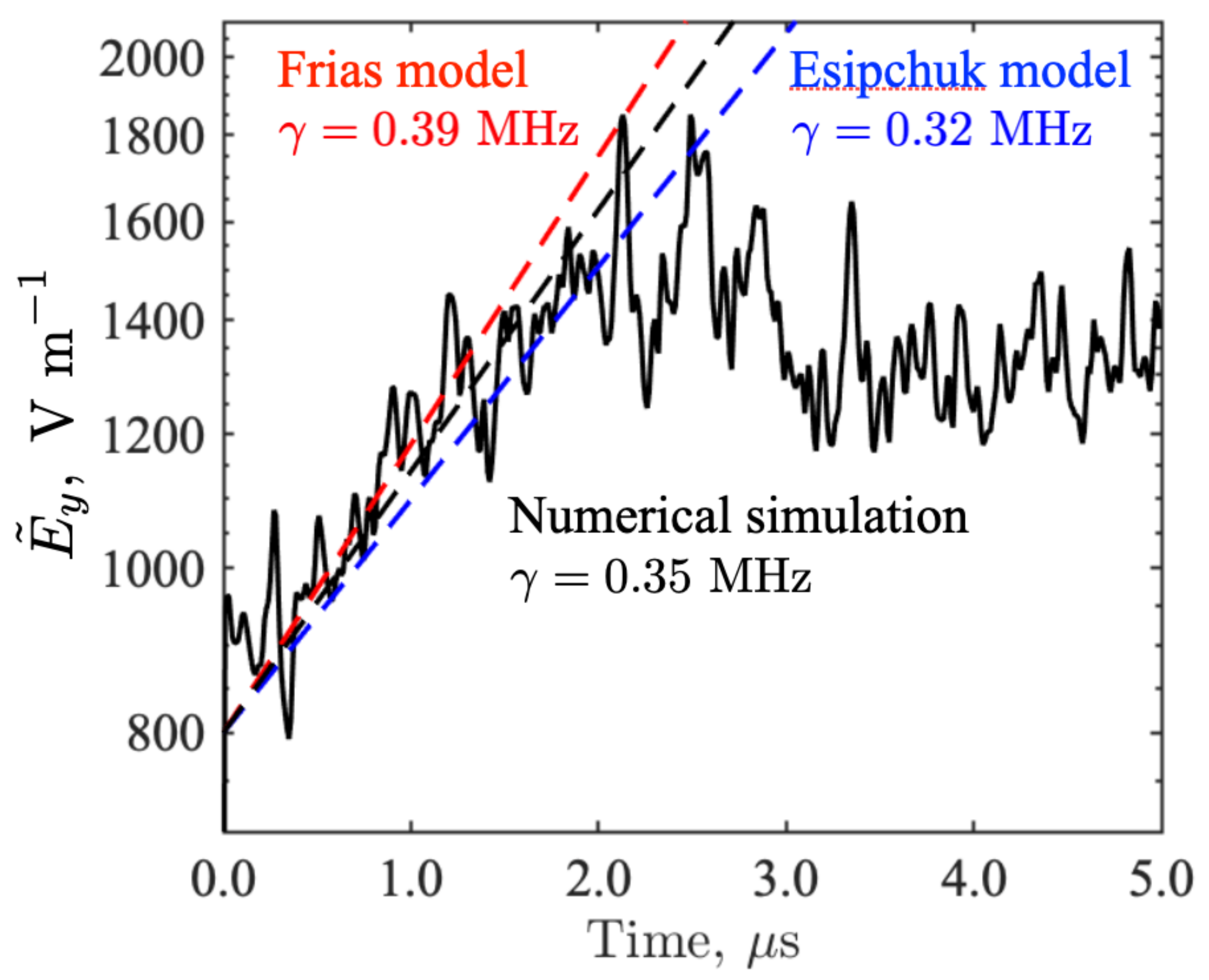}
    \caption{Time history of $\tilde{E}_y$ during the initial evolution of the simulated GDI. The growth rates obtained from the linear GDI models were plotted for comparison.}
    \label{fig:deltaEy}
\end{figure}

The time history of $\tilde{E}_y$ after restarting with an azimuthally uniform profile is shown in Fig. \ref{fig:deltaEy}.
The initial value of $\tilde{E}_y$ is assumed to be caused by the statistical noise on the ion particle model.
Although the time history of $\tilde{E}_y$ is not monotonic, one can recognize that the oscillation grew during the initial phase of 2.0 \textmu s.
The growth rate of the simulated instability was evaluated by fitting a line to the data of $\tilde{E}_y$ in the initial 2.0 \textmu s.
The growth rate of the instability in the numerical simulation was estimated as $\gamma=$ 0.35 MHz.
This value quantitatively agrees with the growth rates obtained from the Esipchuk and Frias perturbation models.
An agreement was confirmed between the present numerical simulation and perturbation models in terms of the growth rate.

The agreement of the wave number, source location, and growth rate of the instability was confirmed between the numerical simulation and perturbation analyses.
This fact demonstrates the relevance between the simulated plasma instability and the theory of GDI.
Here, we simulated one operation condition of one TAL design.
Further detailed comparisons between numerical simulations and perturbation theory for various conditions of SPTs and TALs would extend the validity of the simulation model.

\section{Conclusion}
A hybrid particle-fluid model was developed for the numerical analysis of GDI and cross-field electron transport in Hall thrusters.
The magnetized electron fluid model assumed inertialess electrons and quasineutrality.
The plasma potential was calculated through the electron fluid equations, and DOS was proposed for the stable computation of the potential solver.
The electron energy equation was also stably calculated for the electron temperature using the fully implicit method.
The hybrid model was applied to the axial-azimuthal coordinate of a TAL-type Hall thruster.

The plasma instability was generated in the acceleration and plume regions.
Large electric fields were induced in both the axial and azimuthal directions.
Because it was affected by the electric field, the magnetized electron flow exhibited oscillatory behavior.
The electron flow field showed complex vortex-like structures that consist of waves of various wave numbers.

Owing to plasma instability, the cross-field electron transport was significantly enhanced in the acceleration and plume regions.
In the plume region, the E$\times$B drift caused by the azimuthal electric field was the predominant factor in the induction of the axial electron current.
The effective electron mobility in the acceleration and plume region was greater than the electron mobility given by the classical diffusion theory by one order of magnitude.
The effective Hall parameter was reduced by the instability compared with that expected by the classical diffusion theory.
The simulated GDI weakens the magnetic confinement of the electrons.

The grid convergences of the plasma instability and effective electron mobility are analyzed.
The size of the vortex-like structures in the instability shows a dependence on the grid system.
Small-size vortices were resolved in the fine grid systems.
However, the amplitude of the induced azimuthal electric field and axial electron velocity did not differ significantly depending on the grid resolution.
The grid convergence was confirmed in the effective electron mobility distribution.

The comparison between the simulation results and the perturbation analyses clearly demonstrated that the simulated plasma instability agrees with the theory of GDI.
The FFT analysis of the simulated instability showed that the instability was induced in the acceleration region.
This region is consistent with the locations of high growth rates predicted by the perturbation models.
In addition, the growth rate of the simulated instability was estimated as 0.35 MHz, which quantitatively agreed with those obtained from the perturbation analyses.
The hybrid particle-fluid model was useful for the analyses of the GDI, and the cross-field electron mobility could be significantly enhanced by the GDI.


\begin{acknowledgments}
    This work was supported by JSPS KAKENHI Grant Numbers JP17K14873 and JP20H02346.
\end{acknowledgments}

\section*{Data Availability}
    The numerical data for plotting Figs. \ref{fig:dist1d} \ref{fig:jex1d}, \ref{fig:mueff}, and \ref{fig:Ome_eff} are openly available \cite{KawashimaZenodo2021}.

\appendix





\nocite{*}
\bibliography{reference}

\begin{thebibliography}{49}%
\makeatletter
\providecommand \@ifxundefined [1]{%
 \@ifx{#1\undefined}
}%
\providecommand \@ifnum [1]{%
 \ifnum #1\expandafter \@firstoftwo
 \else \expandafter \@secondoftwo
 \fi
}%
\providecommand \@ifx [1]{%
 \ifx #1\expandafter \@firstoftwo
 \else \expandafter \@secondoftwo
 \fi
}%
\providecommand \natexlab [1]{#1}%
\providecommand \enquote  [1]{``#1''}%
\providecommand \bibnamefont  [1]{#1}%
\providecommand \bibfnamefont [1]{#1}%
\providecommand \citenamefont [1]{#1}%
\providecommand \href@noop [0]{\@secondoftwo}%
\providecommand \href [0]{\begingroup \@sanitize@url \@href}%
\providecommand \@href[1]{\@@startlink{#1}\@@href}%
\providecommand \@@href[1]{\endgroup#1\@@endlink}%
\providecommand \@sanitize@url [0]{\catcode `\\12\catcode `\$12\catcode
  `\&12\catcode `\#12\catcode `\^12\catcode `\_12\catcode `\%12\relax}%
\providecommand \@@startlink[1]{}%
\providecommand \@@endlink[0]{}%
\providecommand \url  [0]{\begingroup\@sanitize@url \@url }%
\providecommand \@url [1]{\endgroup\@href {#1}{\urlprefix }}%
\providecommand \urlprefix  [0]{URL }%
\providecommand \Eprint [0]{\href }%
\providecommand \doibase [0]{http://dx.doi.org/}%
\providecommand \selectlanguage [0]{\@gobble}%
\providecommand \bibinfo  [0]{\@secondoftwo}%
\providecommand \bibfield  [0]{\@secondoftwo}%
\providecommand \translation [1]{[#1]}%
\providecommand \BibitemOpen [0]{}%
\providecommand \bibitemStop [0]{}%
\providecommand \bibitemNoStop [0]{.\EOS\space}%
\providecommand \EOS [0]{\spacefactor3000\relax}%
\providecommand \BibitemShut  [1]{\csname bibitem#1\endcsname}%
\let\auto@bib@innerbib\@empty
\bibitem [{\citenamefont {Choueiri}(2001{\natexlab{a}})}]{Choueiri:2001aa}%
  \BibitemOpen
  \bibfield  {author} {\bibinfo {author} {\bibfnamefont {E.~Y.}\ \bibnamefont
  {Choueiri}},\ }\bibfield  {title} {\enquote {\bibinfo {title} {Fundamental
  difference between the two {H}all thruster variants},}\ }\href {\doibase
  10.1063/1.1409344} {\bibfield  {journal} {\bibinfo  {journal} {Physics of
  Plasmas}\ }\textbf {\bibinfo {volume} {8}},\ \bibinfo {pages} {5025--5033}
  (\bibinfo {year} {2001}{\natexlab{a}})}\BibitemShut {NoStop}%
\bibitem [{\citenamefont {Hamada}\ \emph {et~al.}(2017)\citenamefont {Hamada},
  \citenamefont {Kawashima}, \citenamefont {Komurasaki}, \citenamefont
  {Yamamoto}, \citenamefont {Tahara},\ and\ \citenamefont
  {Miyasaka}}]{HamadaIEPC2017}%
  \BibitemOpen
  \bibfield  {author} {\bibinfo {author} {\bibfnamefont {Y.}~\bibnamefont
  {Hamada}}, \bibinfo {author} {\bibfnamefont {R.}~\bibnamefont {Kawashima}},
  \bibinfo {author} {\bibfnamefont {K.}~\bibnamefont {Komurasaki}}, \bibinfo
  {author} {\bibfnamefont {N.}~\bibnamefont {Yamamoto}}, \bibinfo {author}
  {\bibfnamefont {H.}~\bibnamefont {Tahara}}, \ and\ \bibinfo {author}
  {\bibfnamefont {T.}~\bibnamefont {Miyasaka}},\ }\href@noop {} {\enquote
  {\bibinfo {title} {Development status of 5 kw class anode-layer type {H}all
  thruster: {RAIJIN}94},}\ }\bibinfo {howpublished} {35th International
  Electric Propulsion Conference, IEPC-2017-412} (\bibinfo {year}
  {2017})\BibitemShut {NoStop}%
\bibitem [{\citenamefont {Boeuf}(2017)}]{Boeuf:2017aa}%
  \BibitemOpen
  \bibfield  {author} {\bibinfo {author} {\bibfnamefont {J.-P.}\ \bibnamefont
  {Boeuf}},\ }\bibfield  {title} {\enquote {\bibinfo {title} {Tutorial: Physics
  and modeling of {H}all thrusters},}\ }\href@noop {} {\bibfield  {journal}
  {\bibinfo  {journal} {Journal of Applied Physics}\ }\textbf {\bibinfo
  {volume} {121}},\ \bibinfo {pages} {011101} (\bibinfo {year}
  {2017})}\BibitemShut {NoStop}%
\bibitem [{\citenamefont {Koo}\ and\ \citenamefont {Boyd}(2006)}]{KooPoP2006}%
  \BibitemOpen
  \bibfield  {author} {\bibinfo {author} {\bibfnamefont {J.~W.}\ \bibnamefont
  {Koo}}\ and\ \bibinfo {author} {\bibfnamefont {I.~D.}\ \bibnamefont {Boyd}},\
  }\bibfield  {title} {\enquote {\bibinfo {title} {Modeling of anomalous
  electron mobility in {H}all thrusters},}\ }\href@noop {} {\bibfield
  {journal} {\bibinfo  {journal} {Physics of Plasmas}\ }\textbf {\bibinfo
  {volume} {13}},\ \bibinfo {pages} {033501} (\bibinfo {year}
  {2006})}\BibitemShut {NoStop}%
\bibitem [{\citenamefont {Hagelaar}\ \emph {et~al.}(2002)\citenamefont
  {Hagelaar}, \citenamefont {Bareilles}, \citenamefont {Garrigues},\ and\
  \citenamefont {Boeuf}}]{Hagelaar:2002aa}%
  \BibitemOpen
  \bibfield  {author} {\bibinfo {author} {\bibfnamefont {G.~J.~M.}\
  \bibnamefont {Hagelaar}}, \bibinfo {author} {\bibfnamefont {J.}~\bibnamefont
  {Bareilles}}, \bibinfo {author} {\bibfnamefont {L.}~\bibnamefont
  {Garrigues}}, \ and\ \bibinfo {author} {\bibfnamefont {J.~P.}\ \bibnamefont
  {Boeuf}},\ }\bibfield  {title} {\enquote {\bibinfo {title} {Role of anomalous
  electron transport in a stationary plasma thruster simulation},}\ }\href@noop
  {} {\bibfield  {journal} {\bibinfo  {journal} {Journal of Applied Physics}\
  }\textbf {\bibinfo {volume} {93}},\ \bibinfo {pages} {67--75} (\bibinfo
  {year} {2002})}\BibitemShut {NoStop}%
\bibitem [{\citenamefont {Keidar}\ and\ \citenamefont
  {Beilis}(2006)}]{Keidar:2006aa}%
  \BibitemOpen
  \bibfield  {author} {\bibinfo {author} {\bibfnamefont {M.}~\bibnamefont
  {Keidar}}\ and\ \bibinfo {author} {\bibfnamefont {I.~I.}\ \bibnamefont
  {Beilis}},\ }\bibfield  {title} {\enquote {\bibinfo {title} {Electron
  transport phenomena in plasma devices with {ExB} drift},}\ }\href@noop {}
  {\bibfield  {journal} {\bibinfo  {journal} {IEEE Transactions on Plasma
  Science}\ }\textbf {\bibinfo {volume} {34}},\ \bibinfo {pages} {804--814}
  (\bibinfo {year} {2006})}\BibitemShut {NoStop}%
\bibitem [{\citenamefont {Lucken}\ \emph {et~al.}(2019)\citenamefont {Lucken},
  \citenamefont {Bourdon}, \citenamefont {Lieberman},\ and\ \citenamefont
  {Chabert}}]{Lucken:2019aa}%
  \BibitemOpen
  \bibfield  {author} {\bibinfo {author} {\bibfnamefont {R.}~\bibnamefont
  {Lucken}}, \bibinfo {author} {\bibfnamefont {A.}~\bibnamefont {Bourdon}},
  \bibinfo {author} {\bibfnamefont {M.~A.}\ \bibnamefont {Lieberman}}, \ and\
  \bibinfo {author} {\bibfnamefont {P.}~\bibnamefont {Chabert}},\ }\bibfield
  {title} {\enquote {\bibinfo {title} {Instability-enhanced transport in low
  temperature magnetized plasma},}\ }\href {\doibase 10.1063/1.5094422}
  {\bibfield  {journal} {\bibinfo  {journal} {Physics of Plasmas}\ }\textbf
  {\bibinfo {volume} {26}},\ \bibinfo {pages} {070702} (\bibinfo {year}
  {2019})}\BibitemShut {NoStop}%
\bibitem [{\citenamefont {Hirakawa}\ and\ \citenamefont
  {Arakawa}(1996)}]{Hirakawa:1996aa}%
  \BibitemOpen
  \bibfield  {author} {\bibinfo {author} {\bibfnamefont {M.}~\bibnamefont
  {Hirakawa}}\ and\ \bibinfo {author} {\bibfnamefont {Y.}~\bibnamefont
  {Arakawa}},\ }\href {\doibase doi:10.2514/6.1996-3195} {\enquote {\bibinfo
  {title} {Numerical simulation of plasma particle behavior in a {H}all
  thruster},}\ }\bibinfo {howpublished} {32nd Joint Propulsion Conference and
  Exhibit, AIAA 1996-3195} (\bibinfo {year} {1996})\BibitemShut {NoStop}%
\bibitem [{\citenamefont {Hirakawa}(1997)}]{HirakawaIEPC1997}%
  \BibitemOpen
  \bibfield  {author} {\bibinfo {author} {\bibfnamefont {M.}~\bibnamefont
  {Hirakawa}},\ }\href@noop {} {\enquote {\bibinfo {title} {Electron transport
  mechanism in a {H}all thruster},}\ }\bibinfo {howpublished} {25th
  International Electric Propulsion Conference, IEPC-97-021} (\bibinfo {year}
  {1997})\BibitemShut {NoStop}%
\bibitem [{\citenamefont {Adam}, \citenamefont {H{\'e}ron},\ and\ \citenamefont
  {Laval}(2003)}]{Adam:2003aa}%
  \BibitemOpen
  \bibfield  {author} {\bibinfo {author} {\bibfnamefont {J.~C.}\ \bibnamefont
  {Adam}}, \bibinfo {author} {\bibfnamefont {A.}~\bibnamefont {H{\'e}ron}}, \
  and\ \bibinfo {author} {\bibfnamefont {G.}~\bibnamefont {Laval}},\ }\bibfield
   {title} {\enquote {\bibinfo {title} {Study of stationary plasma thrusters
  using two-dimensional fully kinetic simulations},}\ }\bibfield  {booktitle}
  {\emph {\bibinfo {booktitle} {Physics of Plasmas}},\ }\href {\doibase
  10.1063/1.1632904} {\bibfield  {journal} {\bibinfo  {journal} {Physics of
  Plasmas}\ }\textbf {\bibinfo {volume} {11}},\ \bibinfo {pages} {295--305}
  (\bibinfo {year} {2003})}\BibitemShut {NoStop}%
\bibitem [{\citenamefont {Lafleur}, \citenamefont {Baalrud},\ and\
  \citenamefont {Chabert}(2016{\natexlab{a}})}]{Lafleur:2016aa}%
  \BibitemOpen
  \bibfield  {author} {\bibinfo {author} {\bibfnamefont {T.}~\bibnamefont
  {Lafleur}}, \bibinfo {author} {\bibfnamefont {S.~D.}\ \bibnamefont
  {Baalrud}}, \ and\ \bibinfo {author} {\bibfnamefont {P.}~\bibnamefont
  {Chabert}},\ }\bibfield  {title} {\enquote {\bibinfo {title} {Theory for the
  anomalous electron transport in {H}all effect thrusters. i. insights from
  particle-in-cell simulations},}\ }\href@noop {} {\bibfield  {journal}
  {\bibinfo  {journal} {Physics of Plasmas}\ }\textbf {\bibinfo {volume}
  {23}},\ \bibinfo {pages} {053502} (\bibinfo {year}
  {2016}{\natexlab{a}})}\BibitemShut {NoStop}%
\bibitem [{\citenamefont {Katz}, \citenamefont {Chaplin},\ and\ \citenamefont
  {Lopez~Ortega}(2018)}]{Katz:2018aa}%
  \BibitemOpen
  \bibfield  {author} {\bibinfo {author} {\bibfnamefont {I.}~\bibnamefont
  {Katz}}, \bibinfo {author} {\bibfnamefont {V.~H.}\ \bibnamefont {Chaplin}}, \
  and\ \bibinfo {author} {\bibfnamefont {A.}~\bibnamefont {Lopez~Ortega}},\
  }\bibfield  {title} {\enquote {\bibinfo {title} {Particle-in-cell simulations
  of {H}all thruster acceleration and near plume regions},}\ }\href {\doibase
  10.1063/1.5054009} {\bibfield  {journal} {\bibinfo  {journal} {Physics of
  Plasmas}\ }\textbf {\bibinfo {volume} {25}},\ \bibinfo {pages} {123504}
  (\bibinfo {year} {2018})}\BibitemShut {NoStop}%
\bibitem [{\citenamefont {Lafleur}\ and\ \citenamefont
  {Chabert}(2017)}]{Lafleur:2017aa}%
  \BibitemOpen
  \bibfield  {author} {\bibinfo {author} {\bibfnamefont {T.}~\bibnamefont
  {Lafleur}}\ and\ \bibinfo {author} {\bibfnamefont {P.}~\bibnamefont
  {Chabert}},\ }\bibfield  {title} {\enquote {\bibinfo {title} {The role of
  instability-enhanced friction on `anomalous'electron and ion transport in
  {H}all-effect thrusters},}\ }\href {\doibase 10.1088/1361-6595/aa9efe}
  {\bibfield  {journal} {\bibinfo  {journal} {Plasma Sources Science and
  Technology}\ }\textbf {\bibinfo {volume} {27}},\ \bibinfo {pages} {015003}
  (\bibinfo {year} {2017})}\BibitemShut {NoStop}%
\bibitem [{\citenamefont {Kapulkin}\ and\ \citenamefont
  {Guelman}(2008)}]{Kapulkin:2008aa}%
  \BibitemOpen
  \bibfield  {author} {\bibinfo {author} {\bibfnamefont {A.}~\bibnamefont
  {Kapulkin}}\ and\ \bibinfo {author} {\bibfnamefont {M.~M.}\ \bibnamefont
  {Guelman}},\ }\bibfield  {title} {\enquote {\bibinfo {title} {Low-frequency
  instability in near-anode region of {H}all thruster},}\ }\bibfield
  {booktitle} {\emph {\bibinfo {booktitle} {IEEE Transactions on Plasma
  Science}},\ }\href {\doibase 10.1109/TPS.2008.2003359} {\bibfield  {journal}
  {\bibinfo  {journal} {IEEE Transactions on Plasma Science}\ }\textbf
  {\bibinfo {volume} {36}},\ \bibinfo {pages} {2082--2087} (\bibinfo {year}
  {2008})}\BibitemShut {NoStop}%
\bibitem [{\citenamefont {Simon}(1963)}]{Simon:1963ux}%
  \BibitemOpen
  \bibfield  {author} {\bibinfo {author} {\bibfnamefont {A.}~\bibnamefont
  {Simon}},\ }\bibfield  {title} {\enquote {\bibinfo {title} {Instability of a
  partially ionized plasma in crossed electric and magnetic fields},}\ }\href
  {\doibase 10.1063/1.1706743} {\bibfield  {journal} {\bibinfo  {journal} {The
  Physics of Fluids}\ }\textbf {\bibinfo {volume} {6}},\ \bibinfo {pages}
  {382--388} (\bibinfo {year} {1963})}\BibitemShut {NoStop}%
\bibitem [{\citenamefont {Hoh}(1963)}]{Hoh:1963vi}%
  \BibitemOpen
  \bibfield  {author} {\bibinfo {author} {\bibfnamefont {F.~C.}\ \bibnamefont
  {Hoh}},\ }\bibfield  {title} {\enquote {\bibinfo {title} {Instability of
  penning‐type discharges},}\ }\href {\doibase 10.1063/1.1706878} {\bibfield
  {journal} {\bibinfo  {journal} {The Physics of Fluids}\ }\textbf {\bibinfo
  {volume} {6}},\ \bibinfo {pages} {1184--1191} (\bibinfo {year}
  {1963})}\BibitemShut {NoStop}%
\bibitem [{\citenamefont {Esipchuk}\ and\ \citenamefont
  {Tilinin}(1976)}]{Esipchuk:1976}%
  \BibitemOpen
  \bibfield  {author} {\bibinfo {author} {\bibfnamefont {Y.~B.}\ \bibnamefont
  {Esipchuk}}\ and\ \bibinfo {author} {\bibfnamefont {G.~N.}\ \bibnamefont
  {Tilinin}},\ }\bibfield  {title} {\enquote {\bibinfo {title} {Drift
  instability in a {H}all-current plasma accelerator},}\ }\href@noop {}
  {\bibfield  {journal} {\bibinfo  {journal} {Soviet Physics. Technical
  Physics}\ }\textbf {\bibinfo {volume} {21}},\ \bibinfo {pages} {417}
  (\bibinfo {year} {1976})}\BibitemShut {NoStop}%
\bibitem [{\citenamefont {{Frias}}\ \emph {et~al.}(2012)\citenamefont
  {{Frias}}, \citenamefont {Smolyakov}, \citenamefont {Kaganovich},\ and\
  \citenamefont {Raitses}}]{FriasPoP2012}%
  \BibitemOpen
  \bibfield  {author} {\bibinfo {author} {\bibfnamefont {W.}~\bibnamefont
  {{Frias}}}, \bibinfo {author} {\bibfnamefont {A.~I.}\ \bibnamefont
  {Smolyakov}}, \bibinfo {author} {\bibfnamefont {I.~D.}\ \bibnamefont
  {Kaganovich}}, \ and\ \bibinfo {author} {\bibfnamefont {Y.}~\bibnamefont
  {Raitses}},\ }\bibfield  {title} {\enquote {\bibinfo {title} {Long wavelength
  gradient drift instability in {H}all plasma devices. i. fluid theory},}\
  }\href@noop {} {\bibfield  {journal} {\bibinfo  {journal} {Physics of
  Plasmas}\ }\textbf {\bibinfo {volume} {19}},\ \bibinfo {pages} {072112}
  (\bibinfo {year} {2012})}\BibitemShut {NoStop}%
\bibitem [{\citenamefont {Smolyakov}\ \emph {et~al.}(2016)\citenamefont
  {Smolyakov}, \citenamefont {Chapurin}, \citenamefont {Frias}, \citenamefont
  {Koshkarov}, \citenamefont {Romadanov}, \citenamefont {Tang}, \citenamefont
  {Umansky}, \citenamefont {Raitses}, \citenamefont {Kaganovich},\ and\
  \citenamefont {Lakhin}}]{Smolyakov:2016aa}%
  \BibitemOpen
  \bibfield  {author} {\bibinfo {author} {\bibfnamefont {A.~I.}\ \bibnamefont
  {Smolyakov}}, \bibinfo {author} {\bibfnamefont {O.}~\bibnamefont {Chapurin}},
  \bibinfo {author} {\bibfnamefont {W.}~\bibnamefont {Frias}}, \bibinfo
  {author} {\bibfnamefont {O.}~\bibnamefont {Koshkarov}}, \bibinfo {author}
  {\bibfnamefont {I.}~\bibnamefont {Romadanov}}, \bibinfo {author}
  {\bibfnamefont {T.}~\bibnamefont {Tang}}, \bibinfo {author} {\bibfnamefont
  {M.}~\bibnamefont {Umansky}}, \bibinfo {author} {\bibfnamefont
  {Y.}~\bibnamefont {Raitses}}, \bibinfo {author} {\bibfnamefont {I.~D.}\
  \bibnamefont {Kaganovich}}, \ and\ \bibinfo {author} {\bibfnamefont {V.~P.}\
  \bibnamefont {Lakhin}},\ }\bibfield  {title} {\enquote {\bibinfo {title}
  {Fluid theory and simulations of instabilities, turbulent transport and
  coherent structures in partially-magnetized plasmas of {E}$\times${B}
  discharges},}\ }\href {\doibase 10.1088/0741-3335/59/1/014041} {\bibfield
  {journal} {\bibinfo  {journal} {Plasma Physics and Controlled Fusion}\
  }\textbf {\bibinfo {volume} {59}},\ \bibinfo {pages} {014041} (\bibinfo
  {year} {2016})}\BibitemShut {NoStop}%
\bibitem [{\citenamefont {Escobar}\ and\ \citenamefont
  {Ahedo}(2015)}]{Escobar2015IEPC}%
  \BibitemOpen
  \bibfield  {author} {\bibinfo {author} {\bibfnamefont {D.}~\bibnamefont
  {Escobar}}\ and\ \bibinfo {author} {\bibfnamefont {E.}~\bibnamefont
  {Ahedo}},\ }\href@noop {} {\enquote {\bibinfo {title} {Numerical analysis of
  high-frequency azimuthal oscillations in {H}all thrusters},}\ }\bibinfo
  {howpublished} {34th International Electric Propulsion Conference,
  IEPC-2015-371} (\bibinfo {year} {2015})\BibitemShut {NoStop}%
\bibitem [{\citenamefont {Lam}, \citenamefont {Fernandez},\ and\ \citenamefont
  {Cappelli}(2015)}]{Lam6922570}%
  \BibitemOpen
  \bibfield  {author} {\bibinfo {author} {\bibfnamefont {C.}~\bibnamefont
  {Lam}}, \bibinfo {author} {\bibfnamefont {E.}~\bibnamefont {Fernandez}}, \
  and\ \bibinfo {author} {\bibfnamefont {M.}~\bibnamefont {Cappelli}},\
  }\bibfield  {title} {\enquote {\bibinfo {title} {{A 2-D Hybrid {H}all
  Thruster Simulation That Resolves the ExB Electron Drift Direction}},}\
  }\href@noop {} {\bibfield  {journal} {\bibinfo  {journal} {IEEE Transactions
  on Plasma Science}\ }\textbf {\bibinfo {volume} {43}},\ \bibinfo {pages}
  {86--94} (\bibinfo {year} {2015})}\BibitemShut {NoStop}%
\bibitem [{\citenamefont {Fernandez}, \citenamefont {Dowdy},\ and\
  \citenamefont {Aley}(2015)}]{Fernandez2015iepc}%
  \BibitemOpen
  \bibfield  {author} {\bibinfo {author} {\bibfnamefont {E.}~\bibnamefont
  {Fernandez}}, \bibinfo {author} {\bibfnamefont {C.}~\bibnamefont {Dowdy}}, \
  and\ \bibinfo {author} {\bibfnamefont {J.}~\bibnamefont {Aley}},\ }\href@noop
  {} {\enquote {\bibinfo {title} {Characterization of fluctuations in hybrid
  axial-azimuthal {H}all thruster simulations},}\ }\bibinfo {howpublished}
  {34th International Electric Propulsion Conference, IEPC-2015-313} (\bibinfo
  {year} {2015})\BibitemShut {NoStop}%
\bibitem [{\citenamefont {Hara}\ and\ \citenamefont
  {Boyd}(2015)}]{HaraIEPC2015}%
  \BibitemOpen
  \bibfield  {author} {\bibinfo {author} {\bibfnamefont {K.}~\bibnamefont
  {Hara}}\ and\ \bibinfo {author} {\bibfnamefont {I.~D.}\ \bibnamefont
  {Boyd}},\ }\href@noop {} {\enquote {\bibinfo {title} {Axial-azimuthal
  hybrid-direct kinetic simulation of {H}all effect thrusters},}\ }\bibinfo
  {howpublished} {34th International Electric Propulsion Conference,
  IEPC-2015-286} (\bibinfo {year} {2015})\BibitemShut {NoStop}%
\bibitem [{\citenamefont {Lam}(2015)}]{LamPhDThesis}%
  \BibitemOpen
  \bibfield  {author} {\bibinfo {author} {\bibfnamefont {C.~M.}\ \bibnamefont
  {Lam}},\ }\href@noop {} {\enquote {\bibinfo {title} {Two-dimensional
  axial-azimuthal (z-$\theta$) simulation of cross-field electron transport in
  a {H}all thruster plasma discharge},}\ }\bibinfo {howpublished} {Ph.D.
  Thesis, Stanford University} (\bibinfo {year} {2015})\BibitemShut {NoStop}%
\bibitem [{\citenamefont {Chernyshev}, \citenamefont {Son},\ and\ \citenamefont
  {Gorshkov}(2019)}]{Chernyshev:2019aa}%
  \BibitemOpen
  \bibfield  {author} {\bibinfo {author} {\bibfnamefont {T.}~\bibnamefont
  {Chernyshev}}, \bibinfo {author} {\bibfnamefont {E.}~\bibnamefont {Son}}, \
  and\ \bibinfo {author} {\bibfnamefont {O.}~\bibnamefont {Gorshkov}},\
  }\bibfield  {title} {\enquote {\bibinfo {title} {2d3v kinetic simulation of
  {H}all effect thruster, including azimuthal waves and diamagnetic effect},}\
  }\href {\doibase 10.1088/1361-6463/ab35cb} {\bibfield  {journal} {\bibinfo
  {journal} {Journal of Physics D: Applied Physics}\ }\textbf {\bibinfo
  {volume} {52}},\ \bibinfo {pages} {444002} (\bibinfo {year}
  {2019})}\BibitemShut {NoStop}%
\bibitem [{\citenamefont {Kawashima}, \citenamefont {Hara},\ and\ \citenamefont
  {Komurasaki}(2018)}]{Kawashima:2018ab}%
  \BibitemOpen
  \bibfield  {author} {\bibinfo {author} {\bibfnamefont {R.}~\bibnamefont
  {Kawashima}}, \bibinfo {author} {\bibfnamefont {K.}~\bibnamefont {Hara}}, \
  and\ \bibinfo {author} {\bibfnamefont {K.}~\bibnamefont {Komurasaki}},\
  }\bibfield  {title} {\enquote {\bibinfo {title} {Numerical analysis of
  azimuthal rotating spokes in a crossed-field discharge plasma},}\ }\href
  {\doibase 10.1088/1361-6595/aab39c} {\bibfield  {journal} {\bibinfo
  {journal} {Plasma Sources Science and Technology}\ }\textbf {\bibinfo
  {volume} {27}},\ \bibinfo {pages} {035010} (\bibinfo {year}
  {2018})}\BibitemShut {NoStop}%
\bibitem [{\citenamefont {Fife}(1998)}]{FifeThesis}%
  \BibitemOpen
  \bibfield  {author} {\bibinfo {author} {\bibfnamefont {J.~M.}\ \bibnamefont
  {Fife}},\ }\href@noop {} {\enquote {\bibinfo {title} {Hybrid-{PIC} modeling
  and electrostatic probe survey of {H}all thrusters},}\ }\bibinfo
  {howpublished} {Ph.D. Thesis, Massachusetts Institute of Technology}
  (\bibinfo {year} {1998})\BibitemShut {NoStop}%
\bibitem [{\citenamefont {Komurasaki}\ and\ \citenamefont
  {Arakawa}(1995)}]{Komurasaki:1995fk}%
  \BibitemOpen
  \bibfield  {author} {\bibinfo {author} {\bibfnamefont {K.}~\bibnamefont
  {Komurasaki}}\ and\ \bibinfo {author} {\bibfnamefont {Y.}~\bibnamefont
  {Arakawa}},\ }\bibfield  {title} {\enquote {\bibinfo {title} {Two-dimensional
  numerical model of plasma flow in a {H}all thruster},}\ }\href {\doibase
  10.2514/3.23974} {\bibfield  {journal} {\bibinfo  {journal} {Journal of
  Propulsion and Power}\ }\textbf {\bibinfo {volume} {11}},\ \bibinfo {pages}
  {1317--1323} (\bibinfo {year} {1995})}\BibitemShut {NoStop}%
\bibitem [{\citenamefont {Goebel}\ and\ \citenamefont
  {Katz}(2008)}]{GoebelKatz2008}%
  \BibitemOpen
  \bibfield  {author} {\bibinfo {author} {\bibfnamefont {D.~M.}\ \bibnamefont
  {Goebel}}\ and\ \bibinfo {author} {\bibfnamefont {I.}~\bibnamefont {Katz}},\
  }\href@noop {} {\emph {\bibinfo {title} {Fundamentals of electric propulsion:
  ion and {H}all thrusters}}},\ \bibinfo {edition} {2nd}\ ed.\ (\bibinfo
  {publisher} {John Wiley and Sons},\ \bibinfo {year} {2008})\BibitemShut
  {NoStop}%
\bibitem [{\citenamefont {Greenwood}(2002)}]{Greenwood:2002aa}%
  \BibitemOpen
  \bibfield  {author} {\bibinfo {author} {\bibfnamefont {J.}~\bibnamefont
  {Greenwood}},\ }\bibfield  {title} {\enquote {\bibinfo {title} {The correct
  and incorrect generation of a cosine distribution of scattered particles for
  monte-carlo modelling of vacuum systems},}\ }\href {\doibase
  https://doi.org/10.1016/S0042-207X(02)00173-2} {\bibfield  {journal}
  {\bibinfo  {journal} {Vacuum}\ }\textbf {\bibinfo {volume} {67}},\ \bibinfo
  {pages} {217--222} (\bibinfo {year} {2002})}\BibitemShut {NoStop}%
\bibitem [{\citenamefont {Dugan}\ and\ \citenamefont
  {Sovie}(1967)}]{Dugan_Sovie_1967}%
  \BibitemOpen
  \bibfield  {author} {\bibinfo {author} {\bibfnamefont {J.~J.}\ \bibnamefont
  {Dugan}}\ and\ \bibinfo {author} {\bibfnamefont {R.}~\bibnamefont {Sovie}},\
  }\href@noop {} {\enquote {\bibinfo {title} {Volume ion production costs in
  tenuous plasmas: a general atom theory and detailed results for helium,
  argon, and cesium},}\ }\bibinfo {howpublished} {NASA Technical Note, D-4150}
  (\bibinfo {year} {1967})\BibitemShut {NoStop}%
\bibitem [{\citenamefont {Kawashima}\ \emph {et~al.}(2019)\citenamefont
  {Kawashima}, \citenamefont {Bak}, \citenamefont {Hamada}, \citenamefont
  {Loo}, \citenamefont {Koizumi},\ and\ \citenamefont
  {Komurasaki}}]{KawashimaIEPC2019}%
  \BibitemOpen
  \bibfield  {author} {\bibinfo {author} {\bibfnamefont {R.}~\bibnamefont
  {Kawashima}}, \bibinfo {author} {\bibfnamefont {J.}~\bibnamefont {Bak}},
  \bibinfo {author} {\bibfnamefont {Y.}~\bibnamefont {Hamada}}, \bibinfo
  {author} {\bibfnamefont {B.~V.}\ \bibnamefont {Loo}}, \bibinfo {author}
  {\bibfnamefont {H.}~\bibnamefont {Koizumi}}, \ and\ \bibinfo {author}
  {\bibfnamefont {K.}~\bibnamefont {Komurasaki}},\ }\href@noop {} {\enquote
  {\bibinfo {title} {Coupled simulation of two-dimensional hybrid {H}all
  thruster models},}\ }\bibinfo {howpublished} {36th International Electric
  Propulsion Conference, IEPC-2019-318} (\bibinfo {year} {2019})\BibitemShut
  {NoStop}%
\bibitem [{\citenamefont {Hagelaar}(2007)}]{Hagelaar:2007aa}%
  \BibitemOpen
  \bibfield  {author} {\bibinfo {author} {\bibfnamefont {G.~J.~M.}\
  \bibnamefont {Hagelaar}},\ }\bibfield  {title} {\enquote {\bibinfo {title}
  {Modelling electron transport in magnetized low-temperature discharge
  plasmas},}\ }\href@noop {} {\bibfield  {journal} {\bibinfo  {journal} {Plasma
  Sources Science and Technology}\ }\textbf {\bibinfo {volume} {16}},\ \bibinfo
  {pages} {S57} (\bibinfo {year} {2007})}\BibitemShut {NoStop}%
\bibitem [{\citenamefont {Kawashima}, \citenamefont {Komurasaki},\ and\
  \citenamefont {Sch{\"o}nherr}(2015)}]{Kawashima201559}%
  \BibitemOpen
  \bibfield  {author} {\bibinfo {author} {\bibfnamefont {R.}~\bibnamefont
  {Kawashima}}, \bibinfo {author} {\bibfnamefont {K.}~\bibnamefont
  {Komurasaki}}, \ and\ \bibinfo {author} {\bibfnamefont {T.}~\bibnamefont
  {Sch{\"o}nherr}},\ }\bibfield  {title} {\enquote {\bibinfo {title} {A
  hyperbolic-equation system approach for magnetized electron fluids in
  quasi-neutral plasmas},}\ }\href@noop {} {\bibfield  {journal} {\bibinfo
  {journal} {Journal of Computational Physics}\ }\textbf {\bibinfo {volume}
  {284}},\ \bibinfo {pages} {59--69} (\bibinfo {year} {2015})}\BibitemShut
  {NoStop}%
\bibitem [{\citenamefont {Kawashima}, \citenamefont {Komurasaki},\ and\
  \citenamefont {Sch{\"o}nherr}(2016)}]{Kawashima2016202}%
  \BibitemOpen
  \bibfield  {author} {\bibinfo {author} {\bibfnamefont {R.}~\bibnamefont
  {Kawashima}}, \bibinfo {author} {\bibfnamefont {K.}~\bibnamefont
  {Komurasaki}}, \ and\ \bibinfo {author} {\bibfnamefont {T.}~\bibnamefont
  {Sch{\"o}nherr}},\ }\bibfield  {title} {\enquote {\bibinfo {title} {A
  flux-splitting method for hyperbolic-equation system of magnetized electron
  fluids in quasi-neutral plasmas},}\ }\href@noop {} {\bibfield  {journal}
  {\bibinfo  {journal} {Journal of Computational Physics}\ }\textbf {\bibinfo
  {volume} {310}},\ \bibinfo {pages} {202 -- 212} (\bibinfo {year}
  {2016})}\BibitemShut {NoStop}%
\bibitem [{\citenamefont {Kawashima}\ \emph {et~al.}(2018)\citenamefont
  {Kawashima}, \citenamefont {Wang}, \citenamefont {Chamarthi}, \citenamefont
  {Koizumi},\ and\ \citenamefont {Komurasaki}}]{Kawashima:2018aa}%
  \BibitemOpen
  \bibfield  {author} {\bibinfo {author} {\bibfnamefont {R.}~\bibnamefont
  {Kawashima}}, \bibinfo {author} {\bibfnamefont {Z.}~\bibnamefont {Wang}},
  \bibinfo {author} {\bibfnamefont {A.~S.}\ \bibnamefont {Chamarthi}}, \bibinfo
  {author} {\bibfnamefont {H.}~\bibnamefont {Koizumi}}, \ and\ \bibinfo
  {author} {\bibfnamefont {K.}~\bibnamefont {Komurasaki}},\ }\href@noop {}
  {\enquote {\bibinfo {title} {{Hyperbolic System Approach for Magnetized
  Electron Fluids in ExB Discharge Plasmas}},}\ }\bibinfo {howpublished} {2018
  AIAA Aerospace Sciences Meeting, AIAA 2018-0175} (\bibinfo {year}
  {2018})\BibitemShut {NoStop}%
\bibitem [{\citenamefont {Beam}\ and\ \citenamefont
  {Warming}(1978)}]{Beam:1978aa}%
  \BibitemOpen
  \bibfield  {author} {\bibinfo {author} {\bibfnamefont {R.~M.}\ \bibnamefont
  {Beam}}\ and\ \bibinfo {author} {\bibfnamefont {R.~F.}\ \bibnamefont
  {Warming}},\ }\bibfield  {title} {\enquote {\bibinfo {title} {An implicit
  factored scheme for the compressible navier-stokes equations},}\ }\href
  {\doibase 10.2514/3.60901} {\bibfield  {journal} {\bibinfo  {journal} {AIAA
  Journal}\ }\textbf {\bibinfo {volume} {16}},\ \bibinfo {pages} {393--402}
  (\bibinfo {year} {1978})}\BibitemShut {NoStop}%
\bibitem [{\citenamefont {Bak}\ \emph {et~al.}(2019)\citenamefont {Bak},
  \citenamefont {Kawashima}, \citenamefont {Komurasaki},\ and\ \citenamefont
  {Koizumi}}]{Bak:2019aa}%
  \BibitemOpen
  \bibfield  {author} {\bibinfo {author} {\bibfnamefont {J.}~\bibnamefont
  {Bak}}, \bibinfo {author} {\bibfnamefont {R.}~\bibnamefont {Kawashima}},
  \bibinfo {author} {\bibfnamefont {K.}~\bibnamefont {Komurasaki}}, \ and\
  \bibinfo {author} {\bibfnamefont {H.}~\bibnamefont {Koizumi}},\ }\bibfield
  {title} {\enquote {\bibinfo {title} {Plasma formation and cross-field
  electron transport induced by azimuthal neutral inhomogeneity in an anode
  layer {H}all thruster},}\ }\href {\doibase 10.1063/1.5090931} {\bibfield
  {journal} {\bibinfo  {journal} {Physics of Plasmas}\ }\textbf {\bibinfo
  {volume} {26}},\ \bibinfo {pages} {073505} (\bibinfo {year}
  {2019})}\BibitemShut {NoStop}%
\bibitem [{\citenamefont {Choueiri}(2001{\natexlab{b}})}]{Choueiri:2001ab}%
  \BibitemOpen
  \bibfield  {author} {\bibinfo {author} {\bibfnamefont {E.~Y.}\ \bibnamefont
  {Choueiri}},\ }\bibfield  {title} {\enquote {\bibinfo {title} {Plasma
  oscillations in {H}all thrusters},}\ }\bibfield  {booktitle} {\emph {\bibinfo
  {booktitle} {Physics of Plasmas}},\ }\href {\doibase 10.1063/1.1354644}
  {\bibfield  {journal} {\bibinfo  {journal} {Physics of Plasmas}\ }\textbf
  {\bibinfo {volume} {8}},\ \bibinfo {pages} {1411--1426} (\bibinfo {year}
  {2001}{\natexlab{b}})}\BibitemShut {NoStop}%
\bibitem [{\citenamefont {Parra}\ \emph {et~al.}(2006)\citenamefont {Parra},
  \citenamefont {Ahedo}, \citenamefont {Fife},\ and\ \citenamefont
  {Mart{\'\i}nez-S{\'a}nchez}}]{Parra:2006aa}%
  \BibitemOpen
  \bibfield  {author} {\bibinfo {author} {\bibfnamefont {F.~I.}\ \bibnamefont
  {Parra}}, \bibinfo {author} {\bibfnamefont {E.}~\bibnamefont {Ahedo}},
  \bibinfo {author} {\bibfnamefont {J.~M.}\ \bibnamefont {Fife}}, \ and\
  \bibinfo {author} {\bibfnamefont {M.}~\bibnamefont
  {Mart{\'\i}nez-S{\'a}nchez}},\ }\bibfield  {title} {\enquote {\bibinfo
  {title} {A two-dimensional hybrid model of the {H}all thruster discharge},}\
  }\href {\doibase 10.1063/1.2219165} {\bibfield  {journal} {\bibinfo
  {journal} {Journal of Applied Physics}\ }\textbf {\bibinfo {volume} {100}},\
  \bibinfo {pages} {023304} (\bibinfo {year} {2006})}\BibitemShut {NoStop}%
\bibitem [{\citenamefont {Hamada}\ \emph {et~al.}(2021)\citenamefont {Hamada},
  \citenamefont {Kawashima}, \citenamefont {Bak}, \citenamefont {Komurasaki},\
  and\ \citenamefont {Koizumi}}]{Hamada:2021aa}%
  \BibitemOpen
  \bibfield  {author} {\bibinfo {author} {\bibfnamefont {Y.}~\bibnamefont
  {Hamada}}, \bibinfo {author} {\bibfnamefont {R.}~\bibnamefont {Kawashima}},
  \bibinfo {author} {\bibfnamefont {J.}~\bibnamefont {Bak}}, \bibinfo {author}
  {\bibfnamefont {K.}~\bibnamefont {Komurasaki}}, \ and\ \bibinfo {author}
  {\bibfnamefont {H.}~\bibnamefont {Koizumi}},\ }\bibfield  {title} {\enquote
  {\bibinfo {title} {Characterization of acceleration zone shifting in an
  anode-layer-type {H}all thruster {RAIJIN66}},}\ }\href {\doibase
  https://doi.org/10.1016/j.vacuum.2020.110040} {\bibfield  {journal} {\bibinfo
   {journal} {Vacuum}\ }\textbf {\bibinfo {volume} {186}},\ \bibinfo {pages}
  {110040} (\bibinfo {year} {2021})}\BibitemShut {NoStop}%
\bibitem [{\citenamefont {Kawashima}\ and\ \citenamefont
  {Komurasaki}(2021)}]{KawashimaZenodo2021}%
  \BibitemOpen
  \bibfield  {author} {\bibinfo {author} {\bibfnamefont {R.}~\bibnamefont
  {Kawashima}}\ and\ \bibinfo {author} {\bibfnamefont {K.}~\bibnamefont
  {Komurasaki}},\ }\href {\doibase 10.5281/zenodo.4695404} {\enquote {\bibinfo
  {title} {Data from: {Two-dimensional hybrid model of gradient drift
  instability and enhanced electron transport in a Hall thruster}},}\ }\bibinfo
  {howpublished} {Zenodo. http://dx.doi.org/10.5281/zenodo.4695404} (\bibinfo
  {year} {2021})\BibitemShut {NoStop}%
\bibitem [{\citenamefont {Burrell}(1997)}]{Burrell:1997aa}%
  \BibitemOpen
  \bibfield  {author} {\bibinfo {author} {\bibfnamefont {K.~H.}\ \bibnamefont
  {Burrell}},\ }\bibfield  {title} {\enquote {\bibinfo {title} {Effects of e×b
  velocity shear and magnetic shear on turbulence and transport in magnetic
  confinement devices},}\ }\href {\doibase 10.1063/1.872367} {\bibfield
  {journal} {\bibinfo  {journal} {Physics of Plasmas}\ }\textbf {\bibinfo
  {volume} {4}},\ \bibinfo {pages} {1499--1518} (\bibinfo {year}
  {1997})}\BibitemShut {NoStop}%
\bibitem [{\citenamefont {Lafleur}, \citenamefont {Baalrud},\ and\
  \citenamefont {Chabert}(2016{\natexlab{b}})}]{Lafleur:2016ab}%
  \BibitemOpen
  \bibfield  {author} {\bibinfo {author} {\bibfnamefont {T.}~\bibnamefont
  {Lafleur}}, \bibinfo {author} {\bibfnamefont {S.~D.}\ \bibnamefont
  {Baalrud}}, \ and\ \bibinfo {author} {\bibfnamefont {P.}~\bibnamefont
  {Chabert}},\ }\bibfield  {title} {\enquote {\bibinfo {title} {Theory for the
  anomalous electron transport in {H}all effect thrusters. ii. kinetic
  model},}\ }\href@noop {} {\bibfield  {journal} {\bibinfo  {journal} {Physics
  of Plasmas}\ }\textbf {\bibinfo {volume} {23}},\ \bibinfo {pages} {053503}
  (\bibinfo {year} {2016}{\natexlab{b}})}\BibitemShut {NoStop}%
\bibitem [{\citenamefont {Hofer}\ and\ \citenamefont
  {Gallimore}(2006)}]{Hofer:2006aa}%
  \BibitemOpen
  \bibfield  {author} {\bibinfo {author} {\bibfnamefont {R.~R.}\ \bibnamefont
  {Hofer}}\ and\ \bibinfo {author} {\bibfnamefont {A.~D.}\ \bibnamefont
  {Gallimore}},\ }\bibfield  {title} {\enquote {\bibinfo {title} {High-specific
  impulse {H}all thrusters, part 2: Efficiency analysis},}\ }\href {\doibase
  10.2514/1.15954} {\bibfield  {journal} {\bibinfo  {journal} {Journal of
  Propulsion and Power}\ }\textbf {\bibinfo {volume} {22}},\ \bibinfo {pages}
  {732--740} (\bibinfo {year} {2006})}\BibitemShut {NoStop}%
\bibitem [{\citenamefont {Chamarthi}, \citenamefont {Komurasaki},\ and\
  \citenamefont {Kawashima}(2018)}]{Chamarthi:2018aa}%
  \BibitemOpen
  \bibfield  {author} {\bibinfo {author} {\bibfnamefont {A.~S.}\ \bibnamefont
  {Chamarthi}}, \bibinfo {author} {\bibfnamefont {K.}~\bibnamefont
  {Komurasaki}}, \ and\ \bibinfo {author} {\bibfnamefont {R.}~\bibnamefont
  {Kawashima}},\ }\bibfield  {title} {\enquote {\bibinfo {title} {High-order
  upwind and non-oscillatory approach for steady state diffusion,
  advection--diffusion and application to magnetized electrons},}\ }\href
  {\doibase https://doi.org/10.1016/j.jcp.2018.08.018} {\bibfield  {journal}
  {\bibinfo  {journal} {Journal of Computational Physics}\ }\textbf {\bibinfo
  {volume} {374}},\ \bibinfo {pages} {1120--1151} (\bibinfo {year}
  {2018})}\BibitemShut {NoStop}%
\bibitem [{\citenamefont {Sekerak}\ \emph {et~al.}(2016)\citenamefont
  {Sekerak}, \citenamefont {Gallimore}, \citenamefont {Brown}, \citenamefont
  {Hofer},\ and\ \citenamefont {Polk}}]{Sekerak:2016aa}%
  \BibitemOpen
  \bibfield  {author} {\bibinfo {author} {\bibfnamefont {M.~J.}\ \bibnamefont
  {Sekerak}}, \bibinfo {author} {\bibfnamefont {A.~D.}\ \bibnamefont
  {Gallimore}}, \bibinfo {author} {\bibfnamefont {D.~L.}\ \bibnamefont
  {Brown}}, \bibinfo {author} {\bibfnamefont {R.~R.}\ \bibnamefont {Hofer}}, \
  and\ \bibinfo {author} {\bibfnamefont {J.~E.}\ \bibnamefont {Polk}},\
  }\bibfield  {title} {\enquote {\bibinfo {title} {Mode transitions in
  {H}all-effect thrusters induced by variable magnetic field strength},}\
  }\href {\doibase 10.2514/1.B35709} {\bibfield  {journal} {\bibinfo  {journal}
  {Journal of Propulsion and Power}\ }\textbf {\bibinfo {volume} {32}},\
  \bibinfo {pages} {903--917} (\bibinfo {year} {2016})}\BibitemShut {NoStop}%
\bibitem [{\citenamefont {Bak}\ \emph {et~al.}(2020)\citenamefont {Bak},
  \citenamefont {Van~Loo}, \citenamefont {Kawashima},\ and\ \citenamefont
  {Komurasaki}}]{Bak:2020aa}%
  \BibitemOpen
  \bibfield  {author} {\bibinfo {author} {\bibfnamefont {J.}~\bibnamefont
  {Bak}}, \bibinfo {author} {\bibfnamefont {B.}~\bibnamefont {Van~Loo}},
  \bibinfo {author} {\bibfnamefont {R.}~\bibnamefont {Kawashima}}, \ and\
  \bibinfo {author} {\bibfnamefont {K.}~\bibnamefont {Komurasaki}},\ }\bibfield
   {title} {\enquote {\bibinfo {title} {Discharge characteristics and increased
  electron current during azimuthally nonuniform propellant supply in an anode
  layer {H}all thruster},}\ }\href {\doibase 10.1063/1.5144851} {\bibfield
  {journal} {\bibinfo  {journal} {Journal of Applied Physics}\ }\textbf
  {\bibinfo {volume} {128}},\ \bibinfo {pages} {023302} (\bibinfo {year}
  {2020})}\BibitemShut {NoStop}%
\bibitem [{\citenamefont {Marusov}\ \emph {et~al.}(2019)\citenamefont
  {Marusov}, \citenamefont {Sorokina}, \citenamefont {Lakhin}, \citenamefont
  {Ilgisonis},\ and\ \citenamefont {Smolyakov}}]{Marusov}%
  \BibitemOpen
  \bibfield  {author} {\bibinfo {author} {\bibfnamefont {N.~A.}\ \bibnamefont
  {Marusov}}, \bibinfo {author} {\bibfnamefont {E.~A.}\ \bibnamefont
  {Sorokina}}, \bibinfo {author} {\bibfnamefont {V.~P.}\ \bibnamefont
  {Lakhin}}, \bibinfo {author} {\bibfnamefont {V.~I.}\ \bibnamefont
  {Ilgisonis}}, \ and\ \bibinfo {author} {\bibfnamefont {A.~I.}\ \bibnamefont
  {Smolyakov}},\ }\bibfield  {title} {\enquote {\bibinfo {title}
  {Gradient-drift instability applied to {H}all thrusters},}\ }\href {\doibase
  10.1088/1361-6595/aae23d} {\bibfield  {journal} {\bibinfo  {journal} {Plasma
  Sources Science and Technology}\ }\textbf {\bibinfo {volume} {28}},\ \bibinfo
  {pages} {015002} (\bibinfo {year} {2019})}\BibitemShut {NoStop}%
\end{thebibliography}%

\end{document}